\documentclass[11pt]{article}
\usepackage[margin=1in]{geometry}
\usepackage{amsmath,amssymb,amsthm,mathtools}
\usepackage{thmtools,thm-restate}
\makeatletter
\@ifundefined{newcounteralias}{}{%
  \renewcommand\thmt@autorefsetup{%
    \@xa\def\csname\thmt@envname autorefname\@xa\endcsname\@xa{\thmt@thmname}%
  }%
}
\makeatother
\usepackage{microtype}
\usepackage{xcolor}
\usepackage{todonotes}
\usepackage[colorlinks=true,linkcolor=blue,citecolor=green!45!black,urlcolor=blue]{hyperref}
\usepackage[nameinlink,noabbrev,capitalise]{cleveref}
\usepackage{enumitem}
\usepackage[ruled,section]{algorithm}
\usepackage{algpseudocode}
\usepackage{booktabs}
\usepackage{array}
\usepackage{placeins}
\usepackage{pifont}
\usepackage{lmodern}

\newtheorem{lemma}{Lemma}[section]
\newtheorem{theorem}[lemma]{Theorem}
\newtheorem{claim}{Claim}[lemma]
\crefname{claim}{Claim}{Claims}
\Crefname{claim}{Claim}{Claims}
\newtheorem{problem}[lemma]{Problem}
\crefname{problem}{Problem}{Problems}
\Crefname{problem}{Problem}{Problems}
\theoremstyle{definition}
\newtheorem{definition}[lemma]{Definition}

\crefname{invariance}{Invariant}{Invariants}
\Crefname{invariance}{Invariant}{Invariants}
\theoremstyle{plain}

\newcommand{\eps}{\varepsilon}
\renewcommand{\epsilon}{\varepsilon}

\newcommand{\D}{\mathcal{D}}

\renewcommand{\O}{\mathcal{O}}
\renewcommand{\P}{\mathcal{P}}
\newcommand{\Q}{\mathcal{Q}}
\newcommand{\R}{\mathbb{R}}

\newcommand{\otilde}{\widetilde O}
\newcommand{\thetatilde}{\widetilde\Theta}
\newcommand{\defeq}{\stackrel{\mathrm{{\scriptscriptstyle def}}}{=}}

\newcommand{\grad}{\nabla}

\DeclareMathOperator*{\argmin}{arg\,min}

\DeclareMathOperator{\supp}{supp}

\DeclareMathOperator{\Step}{\mathtt{Step}}
\DeclareMathOperator{\Eval}{\mathtt{Eval}}

\newcommand{\vone}{\vec{1}}

\title{Faster Semi-Streaming and Dynamic Approximate Maximum Weight Matchings
in General Graphs: A Unified Framework}
\author{Jiale Chen\\Stanford University}
\date{}

\begin{document}
\maketitle
\begin{abstract}
We study the $(1-\varepsilon)$-approximate maximum weight matching problem
in general graphs with $n$ vertices and integer edge weights in $[1,W]$.
For $W=n^{O(1)}$, we obtain faster algorithms in the
following settings:

\begin{itemize}[leftmargin=*]
\item \textbf{Semi-streaming.}
Deterministic $O(\varepsilon^{-2})$-pass and
$O(\varepsilon^{-2}\log(1/\varepsilon))$-pass algorithms
for unweighted and weighted matching.
These improve the $O(\varepsilon^{-4})$ and $O(\varepsilon^{-39})$ bounds of
Izumi--Kitamura--Yamaguchi (2025) and Huang--Su (PODC~2023), respectively,
and resolve Assadi's (SOSA~2024) open question
on $O(\varepsilon^{-2})$-pass matching algorithms in general graphs.

\item \textbf{Incremental.}
A deterministic algorithm for weighted matching
processing $m$ edge insertions in total time
$O((m\varepsilon^{-3}+n\varepsilon^{-6})\log^3 n)$.
This is the first incremental matching algorithm
for general graphs with $(\log n/\varepsilon)^{O(1)}$ amortized update time,
including the unweighted case.

\item \textbf{Decremental.}
A randomized algorithm for weighted matching
processing all edge deletions from an $m$-edge graph
in total time
$O((m\varepsilon^{-8}+n\varepsilon^{-47})\log^{O(1)}n)$
with high probability.
On sufficiently dense graphs,
this improves the $O((m\varepsilon^{-20}+n\varepsilon^{-41})\log^{O(1)}n)$
bound from Chen--Sidford--Tu (SODA~2025) and
Bernstein--Chen--Dudeja--Langley--Sidford--Tu (SODA~2025).

\item \textbf{Fully dynamic.}
A deterministic reduction from weighted matching in general graphs
to unweighted matching in bipartite graphs
incurring $O(\varepsilon^{-4}\log^{O(1)}(1/\varepsilon))$
multiplicative update-time overhead, independent of $n$.
This improves upon the
$O(\varepsilon^{-756}\log^{O(1)}n)$
overhead of the randomized reduction of Bernstein--Chen (SODA~2026).
\end{itemize}

Underlying these results is a unified framework
that reduces the problem to repeatedly solving two simpler matching subproblems:
a $(1-\varepsilon)$-approximate maximum weight matching
on a sparse subgraph with $O(n/\varepsilon)$ edges,
and a constant approximation on the remaining edges.

The framework builds on dual entropy regularization,
a smoothing technique underlying efficient solvers
for packing linear programs.
To handle the exponentially many odd-set constraints,
we adapt the smoothing technique to regularize only the vertex duals
and develop a new cut partition that preserves odd-set
feasibility during optimization.
\end{abstract}
\newpage
\tableofcontents
\clearpage

\section{Introduction}
\label{sec:introduction}

We study $(1-\eps)$-approximate
maximum weight matching in general graphs.
Let $G=(V,E,w)$ be an $n$-vertex graph
with integer edge weights $w_e\in[1,W]$.\footnote{Throughout this paper,
we assume $W\geq8$ is a power of two bounded by $n^{O(1)}$.}
A \emph{matching} is a set of vertex-disjoint edges.
The \emph{weight} of a matching $M$ is defined by $w(M)\defeq\sum_{e\in M}w_e$.
A \emph{maximum weight matching}, or \emph{MWM}, is a matching with maximum weight.
We denote the weight of an MWM in $G$ by $\mu_w(G)$.
For $\alpha\in(0,1]$, an $\alpha$-MWM is a matching
with weight at least $\alpha\mu_w(G)$.
For the unweighted variant,
we define MCM, $\mu(G)$, and $\alpha$-MCM analogously,
replacing matching weight with cardinality.

Algorithms for $(1-\eps)$-approximate matching have been extensively studied
in settings that differ in how the graph is accessed and stored,
whether it changes over time,
and how computation is organized.
Sequential algorithms minimize the total running time
with full access to the graph~\cite{DuanP10,DuanP14,AhnG14,ZhengH23}.\footnote{
Although we focus on $(1-\eps)$-approximate matching,
we note that bipartite MCM and MWM can be solved exactly
in almost-linear time via flow reductions
\cite{ChenKLPGS22,BrandCLPGSS23,LiWice26,KyngMGS26}.
An OpenAI preprint reports the same bound
for MCM in general graphs~\cite{OpenAI26}.}
Streaming algorithms process an edge stream in few passes
using limited memory
~\cite{McGregor05,AhnGuha11a,EggertKMS12,AhnG13,AhnG18,Tirodkar18,
GamlathKMS19,AssadiLT21,AssadiJJST22,FischerMU22,AssadiBKL23,HuangS23,
Assadi24,MitrovicMSS25,IzumiKY25}.
Dynamic algorithms maintain a matching
with little work per edge insertion or deletion
\cite{GuptaP13,Gupta14,GrandoniLSSS19,BernsteinGS20,BernsteinDL21,
AssadiBD22,JambulapatiJST22,BlikstadK23,AssadiBKL23,BehnezhadG24,
Liu24,Dudeja23,ChenST25,BernsteinCDLST25,AssadiKK25,
MitrovicS25,BernsteinC26,Chen26}.
Parallel and distributed algorithms divide the work across processors
while keeping the number of computation and communication rounds small
~\cite{GamlathKMS19,AssadiLT21,FischerMU22,HuangS23,
LiuKK23,MitrovicMSS25,MitrovicS25}.

In this paper, we
obtain improved algorithms
in the semi-streaming and dynamic settings.
Below, we define these models and present our results
in comparison with prior work.

\paragraph{Semi-streaming matching.}
In the semi-streaming model,
an algorithm makes passes over an arbitrary-order stream of edges
using $O(n\log^{O(1)}n)$ bits of space for each fixed $\eps>0$.
The goal is to compute an approximate matching in as few passes as possible.

When the pass bound may depend on $\log n$,
a linear dependence on $1/\eps$ is achievable.
In general graphs, randomized algorithms compute a $(1-\eps)$-MWM
in $O(\eps^{-1}\log n/\log\log n)$ passes~\cite{AhnG18,Assadi24}.
The same pass bound is attained deterministically
for $(1-\eps)$-MCM~\cite{Assadi24}.

We focus instead on pass bounds that depend only on $\eps$.
In this setting, a gap between bipartite and general graphs has persisted.
In bipartite graphs, $O(\eps^{-2})$ passes suffice deterministically
for both $(1-\eps)$-MCM and $(1-\eps)$-MWM
~\cite{AssadiLT21,BernsteinCDLST25}.

In general graphs, \cite{FischerMU22} gave the first
$\eps^{-O(1)}$-pass algorithm for $(1-\eps)$-MCM.
Subsequent improvements by \cite{MitrovicMSS25,IzumiKY25}
reduced the pass bound to $O(\eps^{-4})$.
For $(1-\eps)$-MWM, \cite{HuangS23} gave
an $\eps^{-O(1)}$-pass algorithm,
with an $O(\eps^{-39})$ bound recorded
in~\cite[Table~2]{IzumiKY25}.

Whether $O(\eps^{-2})$ passes
suffice for $(1-\eps)$-approximate matching in general graphs,
in both the unweighted and weighted cases,
was posed explicitly in~\cite[Open Question~2]{Assadi24}.
We resolve this question for unweighted matching
and up to a $\log(1/\eps)$ factor for weighted matching.
\begin{restatable}{theorem}{streamingtheorem}
\label{thm:streaming}
There are deterministic semi-streaming algorithms for general graphs
with the following guarantees.
\begin{enumerate}[label=\arabic*.,nosep]
    \item $(1-\eps)$-MCM using $O(\eps^{-2})$ passes
    and $O(\mu(G)\eps^{-2}\log n)$ bits of space.
    \item $(1-\eps)$-MWM using $O(\eps^{-2}\log(1/\eps))$ passes
    and $O(\mu(G)\eps^{-2}\log^2 n)$ bits of space.
\end{enumerate}
\end{restatable}

\paragraph{Dynamic matching.}
In dynamic matching, the graph undergoes adversarial edge updates,
and an algorithm must maintain an explicit $(1-\eps)$-MWM after each update.
The ultimate goal is $(\log n/\eps)^{O(1)}$
 amortized update time~\cite{AssadiBD22,Liu24}.
The fully dynamic model allows both edge insertions and deletions,
while the partially dynamic models
allow only insertions (incremental) or only deletions (decremental).

For incremental matching,
$(\log n/\eps)^{O(1)}$ update times are known in bipartite graphs for
both unweighted and weighted matching~\cite{Gupta14},
with subsequent improvements to $\eps^{-O(1)}$
update time in~\cite{BlikstadK23,BernsteinCDLST25,Chen26}.
For general graphs, however, this goal remained open
even for unweighted matching,
where \cite{GrandoniLSSS19} achieves
$\eps^{-O(1/\eps)}$ amortized update time.
We close this gap for both unweighted and weighted matching.

\begin{restatable}{theorem}{incrementaltheorem}
\label{thm:incremental}
For $0<\eps\leq1/4$, starting from the empty $n$-vertex graph,
an explicit $(1-\eps)$-MWM can be maintained deterministically
through $m$ edge insertions in total time
$O((m\eps^{-3}+n\eps^{-6})\log^3 n)$.
\end{restatable}

For decremental weighted matching in general graphs,
combining the fractional algorithm of \cite{ChenST25}
with the rounding algorithm of \cite{BernsteinCDLST25}
gives total time $O((m\eps^{-20}+n\eps^{-41})\log^{O(1)}n)$.
We improve the running time when the initial graph is sufficiently dense.

\begin{restatable}{theorem}{decrementaltheorem}
\label{thm:decremental}
For $0<\eps\leq1/4$, starting from an $n$-vertex $m$-edge graph,
an explicit $(1-\eps)$-MWM can be maintained with high probability
through edge deletions in total time
$O((m\eps^{-8}+n\eps^{-47})\log^{O(1)}n)$.
\end{restatable}

In the fully dynamic setting,
recent work has made substantial progress
on unweighted matching~\cite{AssadiBKL23,Liu24,BehnezhadG24,AssadiKK25,MitrovicS25}.
A challenge for weighted matching is to carry this and future progress over
with little approximation loss and update-time overhead. A central line of work provides reductions from weighted to unweighted matching.

Early reductions accept an $\alpha$-MCM algorithm
for any $\alpha\in(0,1]$,
independently of the reduction parameter $\eps$,
and convert it to a $(1/2-\eps)\alpha$-MWM in general graphs
with $O(\eps^{-2}\log^2 W)$ update-time overhead~\cite{StubbsW17},
or a $(1-\eps)\alpha$-MWM in bipartite graphs
with $\eps^{-O(1/\eps)}\log W$ overhead~\cite{BernsteinDL21}.

Recent low-loss reductions instead target $(1-\eps)$-MWM
using a $(1-\rho)$-MCM algorithm,
where $\rho$ depends on $\eps$.
In bipartite graphs, matching composition,
introduced in \cite{BernsteinCDLST25} and improved in \cite{Chen26},
gives a
deterministic reduction with multiplicative and additive update-time overheads
of $\otilde(\eps^{-2})$ and $\otilde(\eps^{-4})$,
respectively, using $\rho=\thetatilde(\eps)$.\footnote{
Throughout this paper, $\otilde$, $\thetatilde$, and $\widetilde\Omega$
hide only $\log^{O(1)}(1/\eps)$ factors.
All dependence on $\log n$ is displayed explicitly.}
For general graphs, \cite{BernsteinC26} combines matching composition
with a new primal-dual framework
to obtain a randomized low-loss reduction from general MWM to bipartite MCM
with $\otilde(\eps^{-756}\log^4 n)$ multiplicative overhead,
using $\rho=\thetatilde(\eps^{510})$.

Our reduction improves the multiplicative overhead of \cite{BernsteinC26}
to $\otilde(\eps^{-4})$,
and relaxes the required unweighted accuracy to
$\rho=\thetatilde(\eps)$.
Additionally, our reduction preserves determinism.

\begin{restatable}{theorem}{dynamicreductiontheorem}
\label{thm:dynamic-reduction}
Suppose that, for every $0<\rho\leq1/2$,
there is a fully dynamic algorithm that,
on a bipartite graph with at most $n$ vertices and $m$ edges,
explicitly maintains a $(1-\rho)$-MCM against adaptive adversaries
in amortized update time $U_b(n,m,\rho)$.
Then, for every $0<\eps\leq1/4$,
there is a fully dynamic algorithm that,
on an initially empty general graph with at most $n$ vertices and $m$ edges
and weights in $[1,W]$,
explicitly maintains a $(1-\eps)$-MWM in amortized update time
\[
    \otilde\!\left(\eps^{-4}\right)\cdot
        U_b\!\left(\otilde(n\eps^{-2}),
            \otilde((m+n)\eps^{-2}),
            \thetatilde(\eps)\right).
\]
The reduction preserves determinism.
\end{restatable}

\subsection{Overview of our approach}
\label{sec:intro-approach}

Our results follow from a unified framework
for computing a fractional $(1-\eps)$-MWM.
The framework builds on dual entropy regularization,
a smoothing technique used in efficient solvers
for \emph{packing linear programs} of the form
$\max\{w^\top x:Ax\leq\vone,\ x\geq0\}$ with $A,w\geq0$
\cite{Allen-ZhuO15Parallel,WangMMR15,Allen-ZhuO19}.
Directly smoothing the packing formulation of general matching, however,
would be costly because of the exponentially many odd-set constraints.
We instead regularize only the vertex duals and
develop a new cut partition that preserves odd-set feasibility
during optimization. Together, these ingredients reduce each
iteration to computing a near-optimal matching on a sparse
subgraph with $O(n/\eps)$ edges and a constant-approximate matching
on the remaining edges. This reduction enables our improved
semi-streaming and dynamic algorithms.

For the rest of this overview,
we first introduce the matching linear program
and explain, as background, how dual entropy regularization
works for general packing linear programs.
We then describe the challenges posed by general matching
and how our framework addresses them.
Finally, we compare our framework with previous matching methods
and outline its implementations
in the semi-streaming and dynamic settings.

\paragraph{Matching linear program.}
The matching linear program is $\max_{x\in\P_G}w^\top x$,
where the \emph{matching polytope} $\P_G$ is the convex hull
of the indicator vectors of all matchings in $G$.
In bipartite graphs, $\P_G$ is described by nonnegativity
and the \emph{degree constraints}
$x_v\defeq\sum_{e\ni v}x_e\leq1$ for every $v\in V$.
In general graphs, it additionally needs the \emph{odd-set constraints}
$x(S)\defeq\sum_{e\in E[S]}x_e\leq(|S|-1)/2$
for every $S\in\O_G$,
where $\O_G$ is the family of odd vertex sets of $G$
and $E[S]$ contains the edges with both endpoints in $S$.
The points in $\P_G$ are called \emph{fractional matchings}.
By definition, every fractional matching $x$
can be rounded to an integral one on its support
with weight at least $w^\top x$.

\paragraph{Dual entropy regularization.}
Consider a packing linear program
$\max\{w^\top x:Ax\leq\vone,\ x\geq0\}$ with $A,w\geq0$.
In the Lagrangian formulation of it,
nonnegative dual variables $y$ enforce the row constraints
through the term $y^\top(Ax-\vone)$.
Dual entropy regularization adds a concave generalized entropy $r(y)$
to the maximization over $y$, giving
\[
    \min_{x\geq0}\left\{-w^\top x+
    \max_{y\geq0}\bigl[y^\top(Ax-\vone)+t\cdot r(y)\bigr]\right\},
\]
for $t>0$.
Consequently, the inner maximization yields a differentiable convex objective
with exponential penalties on the row constraints.
To approximately minimize
such objectives,
\cite{Allen-ZhuO15Parallel,WangMMR15} use
multiplicative updates based on truncated gradients,
while \cite{Allen-ZhuO19}
uses randomized coordinate updates that couple gradient and mirror descent.

\paragraph{Challenge 1: Exponentially many constraints.}
For general matching, applying the regularization to every constraint
would introduce exponentially many penalty terms.
We regularize only the vertex duals,
obtaining the objective $F_t$ with exponential penalties only
on the vertex degrees.

Since we do not regularize the odd-set duals,
we must keep the odd-set constraints satisfied
during optimization.
To improve the current fractional matching $x$,
we therefore look for the most negative first-order change
among all feasible directions:
\[
    \min_{d:\,x+d\in\P_G}\grad F_t(x)^\top d.
\]
A direction with negative first-order change
and a sufficiently small step size preserves feasibility and decreases $F_t$.
On the other hand, if no such direction is found,
convexity certifies that $x$ is near-optimal for $F_t$.
We show that constant-approximations to this first-order problem
suffice to compute a $(1-\eps)$-approximate solution
to the matching linear program
in $O(\eps^{-2}\log W)$ iterations.

\paragraph{Challenge 2: Efficient steps preserving odd-set feasibility.}
The next question is how to implement the first-order steps efficiently.
For any MWM $M^*$ under weights $-\grad F_t(x)$,
the direction $\vone_{M^*}-x$ achieves the optimal first-order decrease.
However, computing a constant-approximate MWM $M$ under the same weights
and taking the direction $\vone_M-x$ need not capture a constant fraction
of the optimal first-order decrease from $\vone_{M^*}-x$.

One could instead consider two directions separately:
$-x$, which moves the current fractional matching towards zero,
and $\vone_M$, which increases the coordinates on $M$ equally.
The better direction guarantees a constant fraction of the optimal first-order decrease.
Feasibility, however, is the new concern, especially for
the matching direction $\vone_M$.
Our exponential penalties on the vertex degrees
ensure that a sufficiently small step
along $\vone_M$ preserves the degree constraints.
In general graphs, however,
it may violate an odd-set constraint.

We enforce odd-set feasibility by constructing a partition $\Q$ of the vertices with the following property:
the edges within its parts form a sparse subgraph $H$ with $O(n/\eps)$ edges,
while any matching in the remaining graph $G\setminus H$
can be added with a small step size without violating odd-set constraints.
Using the partition, we decompose the optimal direction as
\[
    \vone_{M^*}-x
    =(\vone_{M^*\cap H}-x)+\vone_{M^*\setminus H}.
\]
The first-order decreases of these two directions sum to the optimal one.
The decomposition
reduces the iteration to two simpler matching subproblems:
computing a near-optimal matching in the sparse graph $H$
and a constant-approximate matching in $G\setminus H$,
both under weights $-\grad F_t(x)$.

\paragraph{Cut partition.}
We now describe how to construct the partition $\Q$.
We use a variant of the auxiliary graph of
Padberg and Rao~\cite{PadbergR82},
which relates slack in odd-set constraints to cut capacity.
Since our goal is an approximate matching,
we maintain our fractional matching $x$ satisfying $x_v\leq 1-2\delta$ for every vertex $v$ and $\delta>0$.
Our construction gives each edge $e$ capacity $x_e$,
adds a root $r$, and gives each root edge $vr$
capacity $1-\delta-x_v\geq\delta$.
The cut capacity of $S\subseteq V$ is then
$f(S)=(1-\delta)|S|-2x(S)$.

Consider a Gomory--Hu tree of the auxiliary graph,
and let $\Q$ consist of the connected components after deleting $r$.
Each part is a minimum vertex-root cut side
and hence has capacity at most $1-\delta$,
the capacity of a singleton cut.
Since each vertex contributes at least $\delta$ through its root edge,
each part has fewer than $1/\delta$ vertices,
giving $O(n/\delta)$ internal edges.

To see why crossing matching steps preserve odd-set feasibility,
fix an odd set $S$ and choose a part $Q$ with $S\cap Q$ odd.
The minimum-cut property gives $f(Q)\leq f(S\cup Q)$,
which together with submodularity implies $f(S\cap Q)\leq f(S)$.
Substituting the formula for $f$
and applying the odd-set constraint to $S\cap Q$ gives
\[
    x(S)\leq x(S\cap Q)+\frac{1-\delta}{2}|S\setminus Q|
    \leq\frac{|S|-1}{2}-\frac{\delta}{2}|S\setminus Q|.
\]
On the other hand,
a crossing matching has at most $|S\setminus Q|$ edges inside $S$,
because each such edge uses a distinct vertex outside $Q$.
Thus a step of size at most $\delta/2$
preserves every odd-set constraint.

For efficient implementation, we give a specialized algorithm that computes the partition 
without constructing a Gomory--Hu tree, avoiding its subpolynomial overhead.

\paragraph{Remark.}
Our first-order framework extends to general packing linear programs,
in which case each iteration calls a constant-approximation solver
with the same constraints and modified objective coefficients,
and no cut partition is needed.
To our knowledge, this extension does not immediately improve
known bounds for packing linear programs,
so we present the framework in the matching setting.

\paragraph{Comparison with prior methods.}
Several earlier methods also use dual variables in the matching linear program
to iteratively improve the matching quality, and have different
strategies to deal with odd-set constraints.

Primal-dual scaling algorithms \cite{DuanP10,DuanP14}
use duals to identify eligible edges
for augmenting the current integral matching,
and use the augmentations to construct blossoms
and adjust duals.
The blossoms form a laminar family,
so only $O(n)$ odd-set duals need to be maintained at any time.
Implementing the augmentations and blossom operations
in semi-streaming and dynamic models, however,
incurs large polynomial factors in $1/\eps$
\cite{HuangS23,BernsteinC26}.
Our framework instead reduces each iteration to simpler matching problems
that can be solved efficiently in these models.

Multiplicative-weights methods \cite{AhnG14,ChenST25}
assign dual values that depend exponentially on constraint violations.
Each iteration then approximately optimizes the objective penalized
by these duals over a bipartite matching polytope.
They perturb the odd-set constraints
to approximate the odd-set duals using only a small family
of nearly maximally violated constraints.
This approximation, however, requires smaller optimization steps
and more iterations.
Our framework instead regularizes only the vertex duals, and enforces
the odd-set feasibility through a cut partition.

\paragraph{Application.}
For the semi-streaming algorithm (\cref{sec:streaming}),
we show that the maintained fractional matching has sparse support,
so we can store it and construct the cut partition
within the space bound.
We then implement the subproblems using existing matching routines
as black boxes.

For dynamic matching, we use
a common strategy that \emph{rebuilds} a new matching with our framework
when the old one is no longer a $(1-\eps)$-approximation.
The efficiency then depends on:
\begin{enumerate}[label=\arabic*.,ref=\arabic*,nosep]
    \item\label{it:rebuild-frequency} How often do we rebuild?
    \item\label{it:rebuild-trigger} How can we efficiently decide when to rebuild?
    \item\label{it:rebuild-computation} How can we efficiently compute the new matching?
\end{enumerate}
We briefly explain how they are addressed in
incremental, fully dynamic and decremental settings.

For incremental matching (\cref{sec:incremental}),
when the current matching is no longer
a $(1-\eps)$-approximation, $\mu_w(G)$ already increases by
a factor of $1+\Omega(\eps)$, which
naturally bounds the number of rebuilds (item~\ref{it:rebuild-frequency}).
The main challenge is efficient detection of when we rebuild
(item~\ref{it:rebuild-trigger}).
Under our framework,
we use first-order evaluations to look for a sufficiently negative direction.
If none is found,
convexity certifies that $x$ remains a good approximation.
Otherwise,
we resume optimization from $x$ using black-box or simple incremental solvers
for the subproblems (item~\ref{it:rebuild-computation}).

For fully dynamic weighted matching (\cref{sec:dynamic}),
a fixed schedule answers both items~\ref{it:rebuild-frequency}
and~\ref{it:rebuild-trigger}:
compute a $(1-\eps/2)$-MWM and wait for $\Theta(\eps\mu_w(G)/W)$ edge updates
before another rebuild.
The main challenge is how to rebuild efficiently (item~\ref{it:rebuild-computation}).
Under our framework, computing the crossing matching direction requires new ideas.
To see its difficulty, consider the edge weights used by the matching subproblems:
\[
    -(\grad F_t(x))_{uv}
    =w_{uv}-y_{t,u}(x)-y_{t,v}(x),
\]
where $y_{t,v}(x)$ is the vertex dual at $v$.
An optimization update can therefore change the weights
on many edges at once.
We exploit the fact that each vertex-dual change
shifts all incident edge weights equally.
Combining graph unfolding~\cite{KaoLST01,BernsteinDL21}
and dynamic vertex toggles~\cite{BernsteinC26},
we handle these changes as edge updates
in an auxiliary bipartite unweighted graph and maintain
the crossing matching direction using any given dynamic MCM algorithm.

For decremental matching (\cref{sec:decremental}), \cite{ChenST25}
answers items~\ref{it:rebuild-frequency} and~\ref{it:rebuild-trigger}
through primal entropy regularization, where
deletion in the maintained matching
forces a decrease in the regularized optimum.
For item~\ref{it:rebuild-computation}, each rebuild must approximately maximize
a primal entropy-regularized matching objective.
We extend our framework to concave objectives to compute these rebuilds.

\subsection{Related work on optimization for
\texorpdfstring{$(1-\eps)$}{(1-epsilon)}-approximate matching}

Multiplicative weight updates provide a general framework
for approximately solving packing and covering linear programs
~\cite{PlotkinST95,AroraHK12}.
Implementations and refinements give
$(1-\eps)$-approximate algorithms for semi-streaming matching
~\cite{AhnG13,AhnGuha11a,AhnG18,Assadi24}
and dynamic matching in bipartite graphs
~\cite{Gupta14,BhattacharyaKS23,Liu24}.

Auction algorithms reduce approximate bipartite MCM and MWM
to maximal-matching subproblems,
giving semi-streaming and massively parallel algorithms
~\cite{AssadiLT21,LiuKK23}.
\cite{ZhengH23} develops a multiplicative auction algorithm
for approximate bipartite MWM in the sequential setting
and extends it to support vertex deletions on one side
and insertions on the other.

Approximate matching algorithms can also be obtained
by reducing matching to more general optimization problems.
Solvers for packing linear programs compute fractional $(1-\eps)$-MWM
in sequential~\cite{KoufogiannakisY14,Allen-ZhuO19},
parallel~\cite{LubyN93,Allen-ZhuO15Parallel,MahoneyRWZ16},
and dynamic~\cite{BhattacharyaKS23} models.
\cite{AssadiJJST22} reduces bipartite unweighted matching
to a box--simplex game
and combines a semi-streaming box--simplex game solver with rounding
to obtain a semi-streaming algorithm for $(1-\eps)$-MCM.

Primal entropy regularization provides robustness against adversarial edge deletion and gives faster decremental matching
algorithms in bipartite~\cite{JambulapatiJST22} and general graphs~\cite{ChenST25}.

\subsection{Preliminaries}
\label{sec:preliminaries}

\paragraph{Approximation notation.}
A \emph{fractional $\alpha$-MWM} is an $x\in\P_G$
with $w^\top x\geq\alpha\mu_w(G)$.
We also use \emph{$(\alpha,\beta)$-MWM} for a matching of weight
at least $\alpha\mu_w(G)-\beta$, where $\beta\geq0$,
and define the fractional and unweighted variants analogously.

\paragraph{Additional assumption.}
We may assume $\eps=n^{-O(1)}$ without loss of generality.
Since edge weights are integral and $W=n^{O(1)}$,
a sufficiently small approximation error
already guarantees an exact maximum weight matching.

\paragraph{Computational model.}
We use the standard word RAM with $\Theta(\log n)$-bit words.

\paragraph{Borrowed routines.}
We use the following known algorithms.
For the matching routines, we discard nonpositive edges
and assume the remaining weights can be scaled
to positive integers represented with $O(\log n)$ bits.

\begin{lemma}[\cite{DuanP14}]
\label{thm:dp-mwm}
Given a graph with $m$ edges,
a $(1-\eps)$-MWM can be computed deterministically
in $O(m\eps^{-1}\log(1/\eps))$ time
using $O(m\log n)$ bits of space.
\end{lemma}

\begin{lemma}[\cite{GhaffariWajc19}]
\label{lem:gw-mwm}
Given an edge stream of a graph with weights in $[1,W]$,
a $1/4$-MWM can be computed deterministically
in one pass using $O(\mu(G)\log W\log n)$ bits of space.
\end{lemma}

\begin{proof}
Use \cite[Algorithm~1]{GhaffariWajc19}
with its approximation parameter set to $1$;
\cite[Lemma~3.1]{GhaffariWajc19} gives the approximation guarantee.
Each accepted edge at least doubles the positive vertex duals
at its endpoints, and every positive dual lies in $[1/2,W]$.
Thus each vertex is incident to $O(\log W)$ accepted edges.
The endpoints of a maximal matching form a vertex cover
of size $O(\mu(G))$,
so only $O(\mu(G)\log W)$ edges are accepted.
Each edge and vertex record uses $O(\log n)$ bits.
\end{proof}

The algorithm below is stated as a semi-streaming
algorithm in one pass in \cite{FeigenbaumKMSZ05}.
It maintains its output matching with $O(1)$ work per edge,
so applying its guarantee to each prefix
gives the following incremental formulation.
\begin{lemma}[\cite{FeigenbaumKMSZ05}]
\label{lem:incremental-mwm}
There is an incremental algorithm that maintains
an explicit
$1/6$-MWM deterministically in $O(1)$ worst-case update time.
\end{lemma}

We use the following algorithm for maximizing
a separable concave objective over the matching polytope.
Here, a number is polynomially bounded if it is zero
or has magnitude between $n^{-O(1)}$ and $n^{O(1)}$.
\begin{lemma}[{\cite[Lemma~5.9]{ChenST25}}]
\label{lem:cst-concave-matching}
Let $G$ have $n$ vertices and $m$ edges,
and let $\Psi(z)=\sum_{e\in E}\psi_e(z_e)$,
where each $\psi_e$ is concave on $[0,1]$.
Suppose each $\psi_e$ is nonnegative on $[0,1]$, satisfies $\psi_e(0)=0$,
has polynomially bounded values at inverse-polynomial arguments,
and is given with an evaluation oracle
and a maximizer in $[n^{-O(1)},1]$.
For $\eps\geq n^{-1/2}\log^{O(1)}n$,
a randomized algorithm returns $z\in\P_G$ with
$\Psi(z)\geq(1-\eps)\max_{u\in\P_G}\Psi(u)$
with high probability in $O((m\eps^{-6}+n\eps^{-13})\log^{O(1)}n)$ time.
The bound counts each evaluation as one operation.
\end{lemma}

\section{A Unified Framework for General Weighted Matching}
\label{sec:optimization}
In this section, we present our framework
that computes a fractional
$(1-\eps)$-MWM in $O(\eps^{-2}\log W)$ iterations.
For $t>0$, we apply entropy regularization to the vertex duals, giving
\[
\begin{aligned}
    F_t(x)&\defeq-w^\top x+
    \max_{y\geq0}\sum_v\left[(x_v-1)y_v
    -t y_v\left(\log\frac{y_v}{W^4}-\log e\right)\right] \\
    &=-w^\top x+t(\log e)\sum_v y_{t,v}(x),
\end{aligned}
\]
where the maximizing vertex duals are
$y_{t,v}(x)=W^4 2^{-(1-x_v)/t}$.\footnote{
Throughout, $\log$ denotes the base-$2$ logarithm,
and $\ln$ denotes the natural logarithm.}

We minimize $F_t$ over $\P_G$.
Denote by $F_t^*\defeq\min_{x\in\P_G}F_t(x)$ the optimal value.
The following lemma shows that, for $t=O(\eps/\log W)$,
it suffices to minimize $F_t$ to additive error
$O(\eps)\cdot\mu_w(G)$.

\begin{lemma}
\label{lem:potential-bias}
Let $0<\eps\leq1$ and $0<t\leq\eps/(16\log W)$.
If $x\in\P_G$ satisfies
$F_t(x)-F_t^*\leq\eps\mu_w(G)/2$,
then $x$ is a fractional $(1-\eps)$-MWM.
\end{lemma}

\begin{proof}
Consider any MWM $M^*$ and set
$z=(1-4t\log W)\vone_{M^*}\in\P_G$.
By definition,
$w^\top z-w^\top x=F_t(x)-F_t(z)
+t(\log e)\sum_v(y_{t,v}(z)-y_{t,v}(x))$.
The first term is at most $\eps\mu_w(G)/2$,
since $F_t^*\leq F_t(z)$.

Now we bound the second term. If $v$ is matched in $M^*$,
$y_{t,v}(z)=1$ and $y_{t,v}(x)\geq0$.
Otherwise $z_v=0\leq x_v$, so $y_{t,v}(z)\leq y_{t,v}(x)$.
Hence $\sum_v(y_{t,v}(z)-y_{t,v}(x))
\leq2|M^*|\leq2\mu_w(G)$.
Combining these bounds with $w^\top z=(1-4t\log W)\mu_w(G)$ gives
$\mu_w(G)-w^\top x
\leq(t(4\log W+2\log e)+\eps/2)\mu_w(G)\leq\eps\mu_w(G)$,
since $W\geq8$.
\end{proof}

We achieve the required accuracy through phased optimization.
We explain and analyze each phase in \cref{sec:local-optimization},
bound the total number of iterations in \cref{sec:decreasing-t},
and reduce each iteration to two matching subproblems
in \cref{sec:matching-oracle-overview}.

\subsection{Optimization within a phase}
\label{sec:local-optimization}
Within a phase, we fix $0<t\leq1/(32\log W)$,
set the degree slack to $\delta=t\log W$,
and let the step size $\eta$ be the largest power of two
not exceeding $t\delta/128$.
We maintain $x$ in the \emph{truncated matching polytope}
$\P_{G,\delta}\defeq\{x\in\R_{\geq0}^E:
x_v\leq1-2\delta,\ \forall v\in V;
x(S)\leq(|S|-1)/2,\ \forall S\in\O_G\}$.
The degree slack enables the efficient cut-partition construction
in \cref{lem:cut-partition} later.

To make progress at $x$, we consider the first-order problem
$\min_{d:\,x+d\in\P_G}\grad F_t(x)^\top d$.
An immediate feasible direction is $d=-x$, i.e., shrinking $x$,
with first-order change $-\grad F_t(x)^\top x$.
If the first-order change is sufficiently negative,
shrinking decreases the potential.
Otherwise, we can show that the curvature at the current
point is bounded, so any sufficiently negative first-order change
gives a potential decrease.
We denote by $A$ the unsigned incidence matrix
and use the following lemma to make progress in each case.

\begin{lemma}
\label{lem:shrinking-curvature}
Let $x\in\P_{G,\delta}$.
\begin{enumerate}[label=\arabic*.,nosep]
    \item If $-\grad F_t(x)^\top x\leq-\delta\mu_w(G)/8$, then
    $F_t(x-\eta x)\leq F_t(x)-\frac{\eta}{2}\grad F_t(x)^\top x$.
    \item If $-\grad F_t(x)^\top x\geq-2\delta\mu_w(G)$, then
    $F_t(x+\eta d)\leq F_t(x)+\eta\grad F_t(x)^\top d
    +2\eta^2\mu_w(G)/t$
    for any $d$ with $\|Ad\|_\infty\leq1$
    and $\|Ad\|_1\leq4\mu(G)$.\footnote{To see where the condition comes from,
for any $x+d\in\P_G$, $d$ is the difference between two fractional matchings,
so the changes in vertex degree satisfy
$\|Ad\|_\infty\leq1$ and $\|Ad\|_1\leq4\mu(G)$.}
\end{enumerate}
\end{lemma}

\begin{proof}
Differentiating $F_t$ gives $\grad F_t(x)=-w+A^\top y_t(x)$,
so $\grad F_t(x)^\top x=-w^\top x+\sum_vx_vy_{t,v}(x)$.

For the first claim, shrinking changes each dual variable to
$y_{t,v}(x-\eta x)=y_{t,v}(x)2^{-\eta x_v/t}$.
Taylor's theorem and $\eta\leq t\delta/128$ give
$F_t(x-\eta x)-F_t(x)
\leq-\eta\grad F_t(x)^\top x
+\eta\delta\sum_vx_v^2y_{t,v}(x)/256$.
Since $x_v\leq 1$, we have
$\sum_vx_v^2y_{t,v}(x)\leq \sum_vx_vy_{t,v}(x)=\grad F_t(x)^\top x+w^\top x$.
Since $w^\top x\leq\mu_w(G)\leq8\grad F_t(x)^\top x/\delta$,
we have
$F_t(x-\eta x)-F_t(x)
\leq\bigl(-\eta+\frac{\eta\delta}{256}(1+\frac{8}{\delta})\bigr)\grad F_t(x)^\top x
\leq-\frac{\eta}{2}\grad F_t(x)^\top x$.

For the second claim, Taylor's theorem gives, for some $s\in[0,1]$,
\[
    F_t(x+\eta d)-F_t(x)
    =\eta\grad F_t(x)^\top d
    +\frac{\eta^2}{2t\log e}\sum_vy_{t,v}(x+s\eta d)(Ad)_v^2.
\]
Since $\|Ad\|_\infty\leq1$, we have
$y_{t,v}(x+s\eta d)=y_{t,v}(x) 2^{s\eta(Ad)_v/t}
\leq 2^{\eta/t}y_{t,v}(x)$.
We now bound the second-order term.
For $x_v>1/2$, we have
$\sum_{v:\,x_v>1/2}y_{t,v}(x)(Ad)_v^2
\leq2\sum_vx_vy_{t,v}(x)
=2(w^\top x+\grad F_t(x)^\top x)
\leq2(1+2\delta)\mu_w(G)$.
For vertices with $x_v\leq1/2$, the bounds on $t$ and $W$ give
$y_{t,v}(x)\leq W^4 2^{-1/(2t)}<\delta/128$.
Hence
$\sum_{v:\,x_v\leq1/2}y_{t,v}(x)(Ad)_v^2
\leq\delta\|Ad\|_2^2/128\leq\delta\mu_w(G)/32$,
since $\|Ad\|_2^2\leq\|Ad\|_\infty\|Ad\|_1
\leq4\mu(G)\leq4\mu_w(G)$.

Combining these bounds yields
$\sum_vy_{t,v}(x+s\eta d)(Ad)_v^2
\leq 2^{\eta/t}(2(1+2\delta)+\delta/32)\mu_w(G)
\leq4\mu_w(G)$.
Substituting into Taylor's formula gives
$F_t(x+\eta d)-F_t(x)
\leq\eta\grad F_t(x)^\top d+2\eta^2\mu_w(G)/t$.
\end{proof}

The framework allows an \emph{approximate first-order oracle}
to support efficient computation with finite precision.

\begin{definition}
\label{def:first-order-oracle}
Given $x\in\P_{G,\delta}$, $t$, and a direction $d$,
the oracle $\Eval(x,t,d)$ returns $v$ satisfying
\[
    |v-\grad F_t(x)^\top d|
    \leq\delta\mu_w(G)/1024.
\]
\end{definition}

To choose a direction with sufficiently negative first-order change
in case two of \cref{lem:shrinking-curvature},
we use a \emph{step oracle} that approximately optimizes the first-order problem
while preserving feasibility and bounding changes in vertex degree.

\begin{definition}
\label{def:step-oracle}
Given $x\in\P_{G,\delta}$,
the oracle $\Step(x,t)$ returns $d$ satisfying:
\begin{enumerate}[label=\arabic*.,nosep]
    \item $x+\eta d\in\P_{G,\delta}$,
    $\|Ad\|_\infty\leq1$, and
    $\|Ad\|_1\leq4\mu(G)$.
    \item For an absolute constant $c\geq1$
    fixed for the oracle,
    the first-order change satisfies
    \[
        \grad F_t(x)^\top d
        \leq\frac{1}{c}
        \min_{d^\prime:\,x+d^\prime\in\P_G}\grad F_t(x)^\top d^\prime
        +c\delta\mu_w(G).
    \]
\end{enumerate}
\end{definition}

Within each phase, the algorithm repeatedly checks whether
shrinking $x$ gives a sufficiently negative first-order change.
If not, it checks the direction returned by the step oracle.
If neither direction gives sufficient decrease, convexity bounds
the remaining gap in the function value, and the phase ends.
Below is the pseudocode of the algorithm.

\begin{algorithm}[H]
\caption{Optimization for fixed $t$}
\label{alg:fixed-phase}
\begin{algorithmic}[1]
\Require $x\in\P_{G,\delta}$ with $\delta,\eta$ as defined in \cref{sec:local-optimization},
and $\mu_w(G)/4\leq L\leq\mu_w(G)$.
\Loop
    \If{$\Eval(x,t,x)\geq3\delta L/4$}
        \State $d\gets-x$.
    \Else
        \State $d\gets\Step(x,t)$.
        \If{$\Eval(x,t,d)>-\delta L/8$}
            \State \Return $x$.
        \EndIf
    \EndIf
    \State $x\gets x+\eta d$.
\EndLoop
\end{algorithmic}
\end{algorithm}

We call one pass through the loop in \cref{alg:fixed-phase} an \emph{iteration}.
The following lemma bounds the total number of iterations within a phase.

\begin{lemma}
\label{lem:centering}
Starting from $x^{(0)}\in\P_{G,\delta}$
with $F_t(x^{(0)})-F_t^*\leq32c^2\delta\mu_w(G)$,
\cref{alg:fixed-phase} returns a feasible $x\in\P_{G,\delta}$
with $F_t(x)-F_t^*\leq2c^2\delta\mu_w(G)$
after $O(1/(t^2\log W))$ iterations.
\end{lemma}

\begin{proof}
Each update is feasible and decreases $F_t$
by at least $15\eta\delta\mu_w(G)/1024$.
If $\Eval(x,t,x)\geq3\delta L/4$, the first-order oracle gives
$\grad F_t(x)^\top x\geq3\delta L/4-\delta\mu_w(G)/1024
\geq\delta\mu_w(G)/8$.
Shrinking preserves feasibility by convexity,
and \cref{lem:shrinking-curvature} gives a decrease
of at least $\eta\delta\mu_w(G)/16$.

Otherwise, since $\mu_w(G)\leq4L$,
$\grad F_t(x)^\top x\leq \Eval(x,t,x)+\delta\mu_w(G)/1024
<3\delta L/4+\delta L/256<\delta L\leq\delta\mu_w(G)$,
so the curvature bound applies to the step-oracle direction.
For an accepted direction,
$\grad F_t(x)^\top d\leq-\delta L/8+\delta\mu_w(G)/1024
\leq-31\delta\mu_w(G)/1024$.
The step oracle guarantees feasibility and the norm bounds.
Since $\eta\leq t\delta/128$, the curvature bound gives
\[
    F_t(x+\eta d)-F_t(x)
    \leq-\frac{31\eta\delta}{1024}\mu_w(G)
    +\frac{2\eta^2}{t}\mu_w(G)
    \leq-\frac{15\eta\delta}{1024}\mu_w(G).
\]
Dividing the initial gap by this decrease and including the stopping iteration
bounds the number of iterations by
$32768c^2/(15\eta)+1\leq2186c^2/\eta=O(1/(t^2\log W))$.

The phase stops only after rejecting an oracle direction $d$.
The stopping test and first-order oracle give
$\grad F_t(x)^\top d>-\delta L/8-\delta\mu_w(G)/1024
\geq-129\delta\mu_w(G)/1024$.
The step-oracle guarantee therefore implies
$\grad F_t(x)^\top d^\prime
\geq c\grad F_t(x)^\top d-c^2\delta\mu_w(G)
\geq-2c^2\delta\mu_w(G)$
for every $d^\prime$ with $x+d^\prime\in\P_G$.
Taking $d^\prime=x^*-x$ for a minimizer $x^*$ of $F_t$ over $\P_G$
and using convexity gives $F_t(x)-F_t^*\leq2c^2\delta\mu_w(G)$.
\end{proof}

\subsection{Changing phase}
\label{sec:decreasing-t}
When the value gap is small,
we move to the next phase by halving $t$ and retaining $x$.
The following lemma shows that $x$ satisfies the condition
of the initial gap for the next phase.

\begin{lemma}
\label{lem:phase-transition}
Let $x\in\P_{G,\delta}$ satisfy
$F_t(x)-F_t^*\leq2c^2\delta\mu_w(G)$,
and set $t'\defeq t/2$ and $\delta'\defeq t^\prime\log W=\delta/2$.
Then $x\in\P_{G,\delta'}$ and
$F_{t'}(x)-F_{t'}^*\leq32c^2\delta'\mu_w(G)$.
\end{lemma}

\begin{proof}
Feasibility follows from $\P_{G,\delta}\subseteq\P_{G,\delta'}$.
To compare the potential gaps, write
\[
    F_{t'}(x)-F_{t'}^*
    =F_t(x)-F_t^*
    +\bigl[F_{t'}(x)-F_{t'}(0)-F_t(x)+F_t(0)\bigr]
    +\bigl[F_{t'}(0)-F_{t'}^*-F_t(0)+F_t^*\bigr].
\]
The gap at $t$ is at most $2c^2\delta\mu_w(G)$.
It remains to bound the two bracketed terms by $0$ and $5\delta\mu_w(G)$,
respectively, giving a total of
$(2c^2+5)\delta\mu_w(G)\leq32c^2\delta'\mu_w(G)$,
since $c\geq1$.

For the first bracketed term, halving $t$ gives
$y_{t',v}(x)=y_{t,v}(x)^2/W^4$ for all $x\in\P_{G,\delta}$.
Since $0\leq y_{t,v}(0)\leq y_{t,v}(x)\leq W^4$, we have
\[
    y_{t',v}(x)-y_{t',v}(0)
    =\frac{y_{t,v}(x)^2-y_{t,v}(0)^2}{W^4}
    =\bigl(y_{t,v}(x)-y_{t,v}(0)\bigr)
    \frac{y_{t,v}(x)+y_{t,v}(0)}{W^4}
    \leq2\bigl(y_{t,v}(x)-y_{t,v}(0)\bigr).
\]
Multiplying by $t'\log e=(t\log e)/2$ and summing over vertices gives
$F_{t'}(x)-F_{t'}(0)\leq F_t(x)-F_t(0)$.

For the second bracketed term, we compare the improvements over the value at zero.
We show that
$F_{t'}(0)-F_{t'}^*\leq\mu_w(G)$
and $F_t(0)-F_t^*\geq(1-5\delta)\mu_w(G)$.
Since $y_{t',v}(z)\geq y_{t',v}(0)$ for every $z\in\P_G$,
$F_{t'}(z)\geq F_{t'}(0)-w^\top z\geq F_{t'}(0)-\mu_w(G)$.
For the lower bound at $t$,
let $M^*$ be an MWM and set $z=(1-4\delta)\vone_{M^*}$.
Matched vertices have $y_{t,v}(z)=1$,
while unmatched vertices have $y_{t,v}(z)=y_{t,v}(0)$.
Hence $F_t(z)-F_t(0)
\leq-(1-4\delta)\mu_w(G)+2t(\log e)|M^*|
\leq-(1-5\delta)\mu_w(G)$.
It follows that $F_t(0)-F_t^*\geq(1-5\delta)\mu_w(G)$.
\end{proof}

We now give the full algorithm,
starting at $x=0$ and $t=1/(32\log W)$ with
a fixed constant $C\geq\max\{64,8c^2\}$,
where $c$ is the oracle constant in \cref{def:step-oracle}.

\begin{algorithm}[H]
\caption{Phased optimization}
\label{alg:phased-optimization}
\begin{algorithmic}[1]
\Require $0<\eps\leq1$.
\State Obtain $L$ such that $\mu_w(G)/4\leq L\leq \mu_w(G)$.
\State $x\gets0$, $t\gets1/(32\log W)$.
\While{$t\geq\eps/(C\log W)$}
    \State $x\gets$ the output of \cref{alg:fixed-phase} on $(x,t,L)$.
    \State $t\gets t/2$.
\EndWhile
\State \Return $x$.
\end{algorithmic}
\end{algorithm}

\begin{lemma}
\label{lem:phased-optimization}
\cref{alg:phased-optimization}
returns a fractional $(1-\eps)$-MWM
in $O(\log W/\eps^2)$ iterations.
\end{lemma}

\begin{proof}
Monotonicity of $y_{t,v}$ gives $F_t(0)-F_t^*\leq\mu_w(G)$.
Since the first phase has $x=0$ and $\delta=1/32$,
it satisfies the initial-gap condition in \cref{lem:centering}.
\Cref{lem:phase-transition} preserves this condition for each subsequent phase.

Let $t$ be the parameter of the last completed phase.
The loop condition in \cref{alg:phased-optimization} gives
$\eps/(C\log W)\leq t<2\eps/(C\log W)$,
so $\eps/C\leq\delta<2\eps/C$.
Its output satisfies
$F_t(x)-F_t^*\leq2c^2\delta\mu_w(G)\leq\eps\mu_w(G)/2$,
since $C\geq8c^2$.
Since also $t\leq\eps/(16\log W)$,
\cref{lem:potential-bias} gives the claimed approximation.

At parameter $t$, \cref{lem:centering} bounds the number of iterations by
$O(1/(t^2\log W))$.
As $t$ halves between phases,
these bounds sum geometrically and are dominated by the last phase,
where $t=\Theta(\eps/\log W)$.
The total is therefore $O(\log W/\eps^2)$ iterations.
\end{proof}

\subsection{Reduction to two matching problems}
\label{sec:matching-oracle-overview}
We now implement $\Step$ through two approximate matching computations
under edge weights $-\grad F_t(x)$.
For any subgraph $H\subseteq G$ and matching $M$,
we decompose the first-order change as
\[
    \grad F_t(x)^\top(\vone_M-x)
    =\grad F_t(x)^\top(\vone_{M\cap H}-x)
    +\grad F_t(x)^\top\vone_{M\setminus H}.
\]
The subtraction of $x$ in the first term
calls for a more accurate matching in $H$,
whereas a constant approximation suffices on the remaining edges.
We compute a $(1-\delta)$-MWM $I$ in $H$
and an $(\alpha,O(\delta\mu_w(G)))$-MWM $J$ on the remaining edges,
for an absolute constant $\alpha>0$.
These matchings give the candidate directions
$(1-2\delta)\vone_I-x$ and $\vone_J$.
The following lemma shows that the direction
with more negative first-order change
satisfies the approximation guarantee in \cref{def:step-oracle}.

\begin{lemma}
\label{lem:matching-gain-decomposition}
Let $x\in\P_{G,\delta}$, $H\subseteq G$.
Under edge weights $-\grad F_t(x)$, let $I$ be a $(1-\delta)$-MWM in $H$,
and $J$ be an $(\alpha,O(\delta\mu_w(G)))$-MWM in $G\setminus H$,
for an absolute constant $\alpha\in(0,1]$.
For a sufficiently large absolute constant $c\geq1$,
there exists $d\in\{(1-2\delta)\vone_I-x,\vone_J\}$ such that
\[
    \grad F_t(x)^\top d
    \leq\frac{1}{c}
    \min_{d^\prime:\,x+d^\prime\in\P_G}\grad F_t(x)^\top d^\prime
    +c\delta\mu_w(G).
\]
\end{lemma}

\begin{proof}
Consider an MWM $M^*$ under edge weight $-\grad F_t(x)$.
Write $I^*=M^*\cap H$ and $J^*=M^*\setminus H$.
Since $-\grad F_t(x)\leq w$, every matching has weight at most $\mu_w(G)$.
The approximation guarantee for $I$ loses at most $\delta\mu_w(G)$
relative to $I^*$,
and scaling $\vone_I$ by $1-2\delta$ loses at most another $2\delta\mu_w(G)$.
Thus $-\grad F_t(x)^\top(\vone_{I^*}-(1-2\delta)\vone_I)
\leq3\delta\mu_w(G)$.
The guarantee for $J$ gives
$-\grad F_t(x)^\top\vone_{J^*}
\leq-\grad F_t(x)^\top\vone_J/\alpha+O(\delta\mu_w(G))$.
The better direction $d$ between these two gives
$-\grad F_t(x)^\top(\vone_{M^*}-x)
\leq-(1+1/\alpha)\grad F_t(x)^\top d+O(\delta\mu_w(G))$.
Rearranging gives the claimed guarantee.
\end{proof}

It remains to ensure feasibility of the candidate updates.
For the additive update $x\gets x+\eta\vone_J$,
the degree constraints hold for any $H$
if every edge in $J$ has positive weight under $-\grad F_t(x)$.
To see this, if $x_v+\eta>1-2\delta$, then
$y_{t,v}(x)>W^2 2^{-\eta/t}>W$,
and incident edges have negative weights under $-\grad F_t(x)$.
Thus those vertices are unmatched in $J$.

To preserve the odd-set constraints as well,
we choose $H$ using a cut partition $\Q$ of $V$.
We call an edge \emph{internal} if its endpoints lie in the same part,
and \emph{crossing} otherwise.
A matching is internal or crossing if all its edges are of that type.
\Cref{lem:cut-partition} constructs a partition $\Q$
whose internal graph $H=(V,\bigcup_{Q\in\Q}E[Q])$
has $O(\mu(G)/\delta)$ edges
and for which $x+\eta\vone_J$ satisfies every odd-set constraint
for any crossing matching $J$.

\begin{lemma}
\label{lem:matching-directions}
Let $x\in\P_{G,\delta}$, and let $\Q$ be a partition
whose odd-set guarantee in \cref{lem:cut-partition} holds for $x$,
with internal graph $H$.
Under edge weights $-\grad F_t(x)$,
let $I$ be a $(1-\delta)$-MWM in $H$
and $J$ an $(\alpha,O(\delta\mu_w(G)))$-MWM in $G\setminus H$
for an absolute constant $\alpha\in(0,1]$,
containing only positive-weight edges.
A direction $d\in\{(1-2\delta)\vone_I-x,\vone_J\}$
satisfies \cref{def:step-oracle}
for a sufficiently large absolute constant $c$.
\end{lemma}

\begin{proof}
For $d=(1-2\delta)\vone_I-x$,
feasibility of $x+\eta d$ follows by convexity.
For $d=\vone_J$, the positive edge weights ensure the degree bounds
as shown above, and \cref{lem:cut-partition}
ensures all odd-set constraints, since $\eta\leq\delta/4$.
Both directions satisfy $\|d\|_1\leq2\mu(G)$
and $\|Ad\|_\infty\leq1$,
so $\|Ad\|_1\leq2\|d\|_1\leq4\mu(G)$.
Finally, \cref{lem:matching-gain-decomposition}
gives the first-order guarantee in \cref{def:step-oracle}.
\end{proof}

\section{Cut Partition for Odd-Set Feasibility}
\label{sec:cut-partition}
We now construct the partition used in \cref{sec:matching-oracle-overview}.
The following lemma gives its sparsity and odd-set feasibility guarantees,
with bounds on construction time and space.
For the space bounds, we assume that each coordinate $x_e$
and the parameter $\delta$ below
have an $O(\log n)$-bit representation,
which holds in all later applications.

\begin{samepage}
\begin{lemma}
\label{lem:cut-partition}
Given $x\in\P_{G,\delta}$,
a partition $\Q$ of $V$ can be computed in
$O(|\supp(x)|+\mu(G)/\delta^3)$ time
using $O(\mu(G)\delta^{-2}\log n)$ additional bits of space,
such that:
\begin{enumerate}[label=\arabic*.,nosep]
    \item The number of internal edges is $O(\mu(G)/\delta)$.
    \item For every crossing matching $J$, odd set $S\in\O_G$,
    and $0\leq \eta\leq \delta/4$,
    $(x+\eta\vone_J)(S)\leq(|S|-1)/2$.
\end{enumerate}
\end{lemma}
\end{samepage}

In this section,
\Cref{sec:cut-properties} defines the partition and proves its structural properties,
\Cref{sec:cut-construction} computes the cut sides,
and \cref{sec:cut-partition-proof} completes the proof of \cref{lem:cut-partition}.

\subsection{Structural properties}
\label{sec:cut-properties}
For the input $x$ of \cref{lem:cut-partition}, set
$\gamma=(1-\delta)/\lceil16(1-\delta)/\delta^2\rceil$
and $\tilde x_e=\gamma\lfloor x_e/\gamma\rfloor$ for every $e\in E$.
We construct the partition from minimum cuts in an auxiliary rooted graph.
Define the rooted undirected capacitated graph
$G_{\tilde x,\delta}\defeq(V\cup\{r\},E\cup\{vr:v\in V\},\tilde x)$,
where $r$ is a new vertex.
Each original edge $e\in E$ has capacity $\tilde x_e$,
and each root edge $vr$ has capacity
$\tilde x_{vr}\defeq1-\delta-\tilde x_v\geq \delta$.

The auxiliary graph turns internal fractional matching mass into cut capacity.
For $S\subseteq V$, let $f(S)$ be its cut capacity in $G_{\tilde x,\delta}$:
\[
    f(S)=\sum_{e\in E:\,|e\cap S|=1}\tilde x_e
    +\sum_{v\in S}\tilde x_{vr}
    =(1-\delta)|S|-2\tilde x(S).
\]
For each $v\in V$, let $Q_v$ be an \emph{inclusion-minimal}
minimum $v$--$r$ cut side containing $v$ in $G_{\tilde x,\delta}$:
no proper subset of $Q_v$ containing $v$ is also a minimum cut side.
This side is unique because minimum $v$--$r$ cut sides containing $v$
are closed under intersection.
The following laminarity property lets us select disjoint parts
from these cut sides.

\begin{lemma}
\label{lem:cut-laminar}
The family $\{Q_v:v\in V\}$ is laminar:
any two members are nested or disjoint.
\end{lemma}

\begin{proof}
Let $A=Q_u$ and $B=Q_v$.
If $u\in B$, then $A\cap B$ contains $u$ and $A\cup B$ contains $v$.
Since $A$ and $B$ are minimum cut sides for $u$ and $v$,
we have $f(A)\leq f(A\cap B)$ and $f(B)\leq f(A\cup B)$.
Together with submodularity, this gives
\[
    f(A)+f(B)
    \leq f(A\cap B)+f(A\cup B)
    \leq f(A)+f(B).
\]
Thus $A\cap B$ is also a minimum $u$--$r$ cut side,
and inclusion-minimality forces $A\subseteq B$.
Symmetrically, $v\in A$ implies $B\subseteq A$.
Otherwise, $A\setminus B$ and $B\setminus A$ contain $u$ and $v$, respectively.
Since $A$ and $B$ are minimum cut sides,
$f(A)\leq f(A\setminus B)$ and $f(B)\leq f(B\setminus A)$.
The undirected cut inequality
$f(A\setminus B)+f(B\setminus A)\leq f(A)+f(B)$
forces equality, so both differences are minimum cut sides.
Inclusion-minimality then forces $A\cap B=\varnothing$.
\end{proof}

Let $\Q$ consist of the \emph{inclusion-maximal} members
of this laminar family: retain $Q_v$ if no $Q_u$ strictly contains it,
and include each retained set only once.
By \cref{lem:cut-laminar}, distinct inclusion-maximal members are disjoint.
Every vertex $v$ belongs to $Q_v$, which is contained in a maximal member,
so $\Q$ covers $V$ and is a partition.

We first bound the number of internal edges by bounding the sizes of the parts.

\begin{lemma}
\label{lem:internal-edge-bound}
The partition $\Q$ has $O(\mu(G)/\delta)$ internal edges.
\end{lemma}

\begin{proof}
Fix a part $Q=Q_v\in\Q$.
Each vertex in $Q$ contributes at least $\delta$ through its root edge.
Since $Q$ is a minimum $v$--$r$ cut side,
$f(Q)\leq f(\{v\})=1-\delta$.
Thus $\delta|Q|\leq f(Q)\leq 1-\delta$, implying $|Q|<1/\delta$.

Rearranging $f(Q)=(1-\delta)|Q|-2\tilde x(Q)$
and using $f(Q)\leq1-\delta$, we obtain
$\tilde x(Q)\geq(1-\delta)(|Q|-1)/2$.
If $|Q|\geq2$, then $|Q|-1\geq|Q|/2$ and $1-\delta\geq1/2$,
so $\tilde x(Q)\geq|Q|/8$.

By Edmonds's theorem~\cite{Edmonds65}, $\sum_e x_e\leq\mu(G)$.
Summing over disjoint parts therefore bounds their total nonsingleton size
by $8\sum_{e\in E}\tilde x_e\leq8\mu(G)$.
Since $G$ is simple and each part has fewer than $1/\delta$ vertices,
the number of internal edges is $O(\mu(G)/\delta)$.
\end{proof}

We next prove that crossing matching steps preserve odd-set feasibility.
For large odd sets, the degree bound suffices.
For a small odd set $S$, choose a part $Q$ with $S\cap Q$ odd.
The minimum-cut inequality provides slack proportional to $|S\setminus Q|$,
while a crossing matching has at most $|S\setminus Q|$ edges inside $S$.

\begin{lemma}
\label{lem:odd-set-preservation}
For any crossing matching $J$, $S\in\O_G$, and $0\leq \eta\leq \delta/4$,
$(x+\eta\vone_J)(S)\leq(|S|-1)/2$.
\end{lemma}

\begin{proof}
We treat small and large odd sets separately.
If $|S|\leq1/\delta$, consider any part $Q=Q_v$ with $S\cap Q\in\O_G$,
and put $T\defeq S\cap Q$ and $U\defeq S\setminus Q$.
Such a part exists because $|S|$ is odd and $\Q$ partitions $V$.
Since $Q_v$ is a minimum $v$--$r$ cut side and $S\cup Q_v$ contains $v$,
we have $f(Q_v)\leq f(S\cup Q_v)$.
Submodularity yields
$f(T)\leq f(S)+f(Q_v)-f(S\cup Q_v)\leq f(S)$.
Since $0\leq x_e-\tilde x_e<\gamma$
and $E[S]\setminus E[T]$ contains at most $|S||U|$ edges,
the cut identity gives
\[
    \begin{aligned}
    x(S)-x(T)
    &\leq\tilde x(S)-\tilde x(T)+\gamma|S||U|\\
    &\leq\frac{1-\delta}{2}|U|+\gamma|S||U|
    \leq\left(\frac12-\frac\delta4\right)|U|.
    \end{aligned}
\]
The last inequality uses $\gamma\leq\delta^2/16$ and $|S|\leq1/\delta$.
Since $T\in\O_G$, feasibility gives $x(T)\leq(|T|-1)/2$,
and hence $x(S)\leq(|S|-1)/2-\delta|U|/4$.

The crossing matching $J$ has no edge with both endpoints in $T$.
Each edge of $J\cap E[S]$ therefore uses a distinct vertex of $U$,
so $|J\cap E[S]|\leq|U|$.
Since $\eta\leq \delta/4$, we get $(x+\eta\vone_J)(S)\leq(|S|-1)/2$.

For $|S|>1/\delta$, the degree bounds suffice:
each vertex has degree at most $1-2\delta+\eta$,
so $(x+\eta\vone_J)(S)\leq(1-2\delta+\eta)|S|/2<(|S|-1)/2$.
\end{proof}

\begin{lemma}
\label{lem:fractional-crossing}
The odd-set guarantee in \cref{lem:cut-partition}
also holds with $\vone_J$ replaced by any nonnegative crossing vector $p$
with $Ap\leq\vone$.
Moreover, $\|Ap\|_1\leq4\mu(G)$.
\end{lemma}

\begin{proof}
In the proof of \cref{lem:odd-set-preservation},
every crossing edge inside $S$ has an endpoint in $U=S\setminus Q$.
Thus $p(S)\leq\sum_{v\in U}(Ap)_v\leq|U|$,
which gives the same small-set bound.
The large-set bound uses only $Ap\leq\vone$.
For the norm bound, the endpoints $C$ of a maximal matching form a vertex cover,
so $\sum_ep_e\leq\sum_{v\in C}(Ap)_v\leq|C|\leq2\mu(G)$.
Therefore $\|Ap\|_1=2\sum_ep_e\leq4\mu(G)$.
\end{proof}

\subsection{Computing the cut sides}
\label{sec:cut-construction}
Recall that $Q_s$ is the inclusion-minimal minimum $s$--$r$ cut side
containing $s$ in $G_{\tilde x,\delta}$.
To compute it, we use an $s$--$r$ maximum flow algorithm.

\begin{lemma}
\label{lem:flow-cut-side}
For any maximum $s$--$r$ flow in $G_{\tilde x,\delta}\setminus\{sr\}$,
the vertices reachable from $s$ in the directed residual graph form $Q_s$.
\end{lemma}

\begin{proof}
Removing $sr$ preserves all minimum cut sides,
since its capacity contributes equally to every $s$--$r$ cut.
Fix a maximum flow, and let $R$ be the vertices reachable from $s$
in its residual graph.
Since the flow is maximum, $r\notin R$.
No residual arc leaves $R$, so every edge crossing $R$ is saturated outward.
Thus the cut capacity equals the flow value, and $R$ is a minimum cut side.

The capacity of $Q_s$ also equals the flow value,
so every edge crossing $Q_s$ is saturated outward.
Therefore, every residual path from $s$ stays inside $Q_s$,
giving $R\subseteq Q_s$.
The inclusion-minimality of $Q_s$ forces $R=Q_s$.
\end{proof}

For a source $s$ with $\tilde x_s>0$, we use
the distance-directed augmenting-path algorithm DD1
of \cite{AhujaO91}, stated in \cref{alg:source-cut}.
Its labels satisfy $h(r)=0$ and
$h(u)\leq h(v)+1$ for every residual arc $(u,v)$.
A residual arc $(u,v)$ is \emph{admissible} if $h(u)=h(v)+1$.
The algorithm terminates the search if a label
strictly between the source's and the root's is unused,
because a residual path from the source to the root would
have to pass through that label.
Each vertex has a \emph{current adjacency pointer},
which records the next scan of its outgoing arcs.

In the implementation, capacities and flows are stored in units of $\gamma$.
Since $(1-\delta)/\gamma=\lceil16(1-\delta)/\delta^2\rceil$,
they are integers of magnitude $O(\delta^{-2})$.
We store labels and adjacency pointers only for accessed vertices,
and flow values only for accessed edges.
At each vertex other than $s,r$, we inspect its root edge first.

\begin{algorithm}[H]
\caption{Computing maximum $s$--$r$ flow in $G_{\tilde x,\delta}\setminus\{sr\}$}
\label{alg:source-cut}
\small
\begin{algorithmic}[1]
\State Initialize zero flow and the path $P=(s)$.
\State Set $h(r)=0$, $h(s)=2$, and $h(v)=1$ for $v\notin\{s,r\}$.
\State Initialize current adjacency pointers at the start of each list.
\State Set $N_0=N_2=1$, $N_1=|V|-1$, and $N_i=0$ for $i\geq3$, which count vertices with the corresponding label.
\While{$h(s)<|V|+1$}
    \State Let $v$ be the last vertex of $P$.
    \If{$v=r$}
        \State Increase the flow along $P$ by the minimum residual capacity of its arcs.
        \State Set $P=(s)$.
    \ElsIf{an admissible arc $(v,w)$ is found from the current pointer}
        \State Append $(v,w)$ to $P$.
    \Else
        \State Set $k=h(v)$ and $h(v)=1+\min\{h(w):(v,w)\text{ is residual}\}$.
        \Comment{$\min\varnothing=+\infty$}
        \If{$h(s)\geq|V|+1$}
            \State \textbf{break}
        \EndIf
        \State Set $N_k=N_k-1$ and $N_{h(v)}=N_{h(v)}+1$.
        \State Set the current pointer to the first admissible arc.
        \If{$N_k=0$}
            \State \textbf{break}
        \EndIf
        \If{$v\neq s$}
            \State Delete the last arc of $P$.
        \EndIf
    \EndIf
\EndWhile
\State \Return the vertices reachable from $s$ in the final residual graph.
\end{algorithmic}
\end{algorithm}

\begin{lemma}
\label{lem:cut-construction}
Given the original edges $e$ with $\tilde x_e>0$ in adjacency lists,
\cref{alg:source-cut} computes $Q_s$ for any $s$ with $\tilde x_s>0$ in
$O((\tilde x_s/\gamma)\lceil\tilde x_s/\delta\rceil)$ time
using $O((\tilde x_s/\gamma)\log n)$ additional bits of space.
\end{lemma}

\begin{proof}
Correctness follows from \cite{AhujaO91},
and \cref{lem:flow-cut-side} proves that the returned side is $Q_s$.

Let $K$ be the number of augmentations.
All capacities are multiples of $\gamma$,
so each augmentation carries at least $\gamma$.
Since the flow value is at most $\tilde x_s$, we have $K\leq\tilde x_s/\gamma$.

Let $T$ contain the vertices other than $s$ whose root edges become saturated.
If $T=\varnothing$, each augmentation follows $s,v,r$ and saturates $sv$.
Thus the search scans only the at most
$K$ original edges incident to $s$
and the root edges at their other endpoints,
taking $O(K)$ time and $O(K\log n)$ bits of space.

Now assume that $T\neq\varnothing$.
Since flow on root edges never decreases,
each saturated root edge carries at least $\delta$,
and total flow is at most $\tilde x_s$, we have
$|T|\leq\lfloor\tilde x_s/\delta\rfloor$.
During the execution of the algorithm,
only $T\cup\{s\}$ can have labels at least two,
so $h(s)=O(|T|)$ before each iteration.
Consequently, the length of each augmenting path is $O(|T|)$.
Since the label only increases, the number of relabelings
per vertex is also $O(|T|)$.

Let $M$ count accessed vertices and distinct examined edges,
including the final residual search.
Since pointers reset only on relabeling,
unsuccessful pointer advances and relabel scans cost $O(M|T|)$ time.
Each relabeling removes at most one arc from the current path,
so appending and later removing these arcs costs $O(M|T|)$ time.
Every other appended arc is either used in an augmentation
or remains on the final path.
The $K$ augmenting paths each have $O(|T|)$ arcs,
so the corresponding extensions and flow updates cost $O(K|T|)$ time.
At most $M$ arcs remain at termination,
so the remaining extensions cost $O(M)$ time.
Initializing only accessed vertex and edge records uses
$O(M\log n)$ bits of space
and total time $O((M+K)|T|)$, including the final search.

We now bound $M$. During the augmenting-path search,
we inspect $vr$ first whenever the current path
ends at $v\notin\{s,r\}$.
If $vr$ is unsaturated, then $h(v)=1$
and appending $vr$ completes an augmenting path.
Thus original adjacency lists are scanned only at vertices in $T\cup\{s\}$.
Since $G$ is simple, there are $O(|T|^2)$ original edges inside this set.
We therefore only need to count newly examined edges leaving it.

For a relabel scan at $v$ with $h(v)=1$,
the root edge $vr$ is saturated and no arc is admissible.
We skip the admissibility scan and compute the relabel minimum directly.
Since every residual neighbor has label at least one,
we stop this scan at the first residual neighbor of label one,
or scan the whole list if none exists.
Every new edge leaving $T\cup\{s\}$ has zero flow
and reaches a vertex with an unsaturated root edge, whose label is one.
Thus this scan examines at most one new edge leaving $T\cup\{s\}$.
Only vertices in $T$ can be relabeled from label one, each at most once,
giving at most $|T|$ such new edges over all these scans.
For a relabel scan at label at least two,
all entries have already been examined,
so no new edges are examined.

During an admissibility scan, a new edge leaving $T\cup\{s\}$
immediately completes an augmenting path.
All admissibility scans together therefore examine at most $K$ such new edges.

The final residual search cannot reach an original vertex outside $T\cup\{s\}$,
since its unsaturated root edge would give an augmenting path.
A new edge leaving this set has zero flow and would reach such a vertex,
so this search examines no new edge leaving the set.

Including accessed vertices and their root edges changes these counts
by only a constant factor, so $M=O(K+|T|^2)$.
Since $\tilde x_s\leq1$ and $\gamma\leq \delta^2/16$,
the bounds on $K$ and $|T|$ give $M=O(\tilde x_s/\gamma)$,
which proves the space bound.
Substituting this and $|T|\leq\tilde x_s/\delta$
into $O((M+K)|T|)$ gives the claimed time bound.

\end{proof}

\subsection{Constructing the partition}
\label{sec:cut-partition-proof}
Any vertex with $\tilde x_v=0$ belongs to a singleton part $Q_v=\{v\}$.
We represent those singleton parts implicitly.
For each remaining vertex $v$, compute $Q_v$ using \cref{lem:cut-construction}.
Group these sides by size and process the groups from largest to smallest,
starting with all remaining vertices unmarked.
If the source $v$ of $Q_v$ is unmarked, retain $Q_v$ and mark all its vertices.
The retained sides and the implicit singletons form $\Q$.

\begin{proof}[Proof of \cref{lem:cut-partition}]
By \cref{lem:cut-laminar}, intersecting cut sides are nested.
When $Q_v$ is processed, every retained side is at least as large,
so $Q_v$ is either contained in one or disjoint from all of them.
Its source $v$ is marked exactly in the first case,
so the procedure retains precisely the distinct inclusion-maximal sides.
The two properties follow from
\cref{lem:internal-edge-bound,lem:odd-set-preservation}.

The initial scan costs $O(|\supp(x)|)$ time.
Since $\tilde x_v\leq1$, \cref{lem:cut-construction}
bounds each source search by $O(\tilde x_v/(\gamma \delta))$ time.
Since $\sum_v\tilde x_v=2\sum_e\tilde x_e\leq2\mu(G)$,
all source searches together cost $O(\mu(G)/(\gamma \delta))$ time.

For each computed side $Q_v$, the final residual search visits all its vertices,
so the proof of \cref{lem:cut-construction} gives $|Q_v|=O(\tilde x_v/\gamma)$.
Thus all returned sides have total size $O(\mu(G)/\gamma)$.
Grouping them in an array indexed up to the largest side size
and marking their vertices therefore take $O(\mu(G)/\gamma)$ time
and $O((\mu(G)/\gamma)\log n)$ bits of space.

Every stored edge has capacity at least $\gamma$,
so there are $O(\mu(G)/\gamma)$ edge and incident-vertex records.
Each record uses $O(\log n)$ bits for identifiers, pointers,
capacities, degrees, and search state.
Each source search uses $O((\tilde x_v/\gamma)\log n)$ additional bits.
Using the same workspace for all searches keeps the total at
$O((\mu(G)/\gamma)\log n)$ bits.

Substituting $\gamma=\Theta(\delta^2)$ yields
$O(|\supp(x)|+\mu(G)/\delta^3)$ time
and $O(\mu(G)\delta^{-2}\log n)$ additional bits of space.
\end{proof}

\section{Semi-Streaming General Matching}
\label{sec:streaming}
In the semi-streaming model, we make sequential passes
over an arbitrary-order edge stream using $O(n\log^{O(1)} n)$ space
for fixed $\eps$.
We obtain the following bounds in this section.

\begin{samepage}
\streamingtheorem*
\end{samepage}

To implement \cref{alg:phased-optimization,alg:fixed-phase},
we provide both $\Step$ and $\Eval$ in the semi-streaming setting.
The implementation of $\Eval$ below uses stored data
without additional stream passes.
Based on \cref{lem:matching-directions},
the implementation of $\Step$ reduces to constructing the cut partition
$\Q$ in \cref{lem:cut-partition} and computing the two matching directions.
We construct $\Q$ from the current iterate,
collect the sparse internal graph in one pass
and compute the internal matching from the stored edges.
For the crossing matching,
we use a one-pass algorithm to compute a constant approximation.

In the remainder of this section, we give the weight-dependent algorithm
and then remove the dependence using the
weight-reduction framework of \cite{BernsteinCDLST25}.

\subsection{Weight-dependent algorithm}
\label{sec:streaming-bounded}
We first describe the stored state and implement $\Eval$,
then implement $\Step$ and combine both oracles in the phased algorithm.

\subsubsection{First-order evaluations}
\label{sec:streaming-eval}
We store $x$ on its support in adjacency lists
together with the original edge weights and vertex degrees.
For the word-RAM implementation,
round the updated coordinates downward to multiples of $h$,
where $h$ is a sufficiently small inverse-polynomial power of two.
If $z=x+\eta d$ and $x'=h\lfloor z/h\rfloor$,
then $x'$ is feasible and
$F_t(x')-F_t(z)\leq w^\top(z-x')\leq mWh$,
because decreasing coordinates does not increase the vertex duals.
For sufficiently small $h$,
each update retains a constant fraction of its progress,
and the approximation error and asymptotic iteration bounds are preserved.

Let $T=O(\eps^{-2}\log W)$ be the iteration bound
from \cref{lem:phased-optimization} at accuracy $\eps/2$.
Since each update introduces edges from at most one matching,
the following sparsity bound follows directly.

\begin{lemma}
\label{lem:streaming-support}
After $i$ updates, $|\supp(x)|\leq i\mu(G)$.
For at most $T$ updates,
our implementation stores the iterate and its support
using $O(\mu(G)\eps^{-2}\log W\log n)$ bits.
\end{lemma}

We also give a sparse representation of the vertex duals
for efficient first-order evaluations.
Set $\xi=\delta/8192$.
For vertices with $x_v\leq1/4$, set $\widehat y_v=\xi$.
For vertices with $x_v>1/4$, compute and store $\widehat y_v$
as a multiple of $\xi/4$ satisfying
$y_{t,v}(x)\leq\widehat y_v\leq y_{t,v}(x)+\xi$.
Define $\widehat g_{uv}\defeq-w_{uv}+\widehat y_u+\widehat y_v$.

For a queried direction
$d\in\{x,(1-2\delta)\vone_I-x,\vone_J\}$,
write $d=\lambda x+\nu\vone_M$ with $|\lambda|,|\nu|\leq1$.
Given $M$ and its original weights, return
\[
    \Eval(x,t,d)=
    \lambda\sum_{e\in\supp(x)}\widehat g_e x_e
    +\nu\sum_{e\in M}\widehat g_e.
\]

\begin{lemma}
\label{lem:streaming-eval}
The duals satisfy $y_t(x)\leq\widehat y\leq y_t(x)+\xi\vone$,
and $\Eval$ satisfies \cref{def:first-order-oracle}
for every queried direction.
Each evaluation, including computing $\widehat y$,
takes $O(|\supp(x)|+\mu(G)\log(W/\delta))$ time
and uses $O(\mu(G)\log n)$ additional bits
beyond the stored iterate.
\end{lemma}

\begin{proof}
If $x_v\leq1/4$, then
$y_{t,v}(x)\leq W^4 2^{-3/(4t)}<\xi$.
At every other vertex, the dual bounds hold by construction.
Thus $0\leq\widehat y-y_t(x)\leq\xi\vone$.
Each queried direction satisfies $\|Ad\|_1\leq4\mu(G)$, so
\[
    |\Eval(x,t,d)-\grad F_t(x)^\top d|
    =\bigl|(\widehat y-y_t(x))^\top Ad\bigr|
    \leq4\xi\mu(G)\leq\delta\mu_w(G)/1024.
\]

Since $\sum_vx_v\leq2\mu(G)$,
at most $8\mu(G)$ vertices require numerical dual evaluations.
Fixed-point arithmetic computes each of these duals
to the prescribed accuracy in $O(\log(W/\delta))$ word operations,
using $O(\log n)$ working bits.
Scanning the stored adjacency lists computes the vertex degrees
in $O(|\supp(x)|)$ time,
and evaluating the two sums takes $O(|\supp(x)|+|M|)$ time.
The $O(\mu(G))$ explicit dual entries and the queried matching
use $O(\mu(G)\log n)$ additional bits.
\end{proof}

\subsubsection{Step oracle}
\label{sec:streaming-step}
The step oracle applies \cref{lem:cut-partition} to $x$
and computes the two matchings in \cref{lem:matching-directions},
using the approximate duals from \cref{sec:streaming-eval}.

\begin{algorithm}[H]
\caption{Semi-streaming $\Step(x,t)$}
\label{alg:streaming-step}
\small
\begin{algorithmic}[1]
\Require $x\in\P_{G,\delta}$, $\delta=t\log W$,
    and the stored approximate duals $\widehat y$
    from \cref{sec:streaming-eval}.
\State Apply \cref{lem:cut-partition} to $x$ to obtain $\Q$.
\State Set $\xi=\delta/8192$ and
    $\widehat r_{uv}=\xi\lfloor\max\{0,w_{uv}-\widehat y_u-\widehat y_v\}/\xi\rfloor$.
\State Collect $H=(V,\bigcup_{Q\in\Q}E[Q])$.
\State $I\gets$ a $(1-\delta)$-MWM in $H$ under weights $\widehat r$,
    using \cref{thm:dp-mwm}.
\State $J\gets$ a $1/4$-MWM in $G\setminus H$ under weights $\widehat r$,
    using \cref{lem:gw-mwm}.
\State \Return $\argmin_{d\in\{(1-2\delta)\vone_I-x,\vone_J\}}
    \Eval(x,t,d)$.
\end{algorithmic}
\end{algorithm}

\begin{lemma}
\label{lem:streaming-step}
\Cref{alg:streaming-step} satisfies \cref{def:step-oracle}
using $O(1)$ passes and
$O(\mu(G)(\delta^{-2}+\log W)\log n)$ additional bits of space.
\end{lemma}

\begin{proof}
\Cref{lem:cut-partition}, applied to $x$,
gives the odd-set guarantee for the crossing direction.

The duals from \cref{lem:streaming-eval} give
$0\leq\max\{0,-\grad F_t(x)_e\}-\widehat r_e\leq3\xi$.
Replacing $\max\{0,-\grad F_t(x)_e\}$ by $\widehat r_e$
changes the weight of any matching by at most $3\xi\mu(G)$.
Both matchings contain only edges with $\widehat r_e>0$,
so $-\grad F_t(x)_e>0$ on these edges.
Choosing the direction using $\Eval$ loses at most
$\delta\mu_w(G)/512$ in first-order change.
The proof of \cref{lem:matching-directions} absorbs these errors
into the absolute constant $c$ in \cref{def:step-oracle}.

The cut construction scans the stored support and uses
$O(\mu(G)\delta^{-2}\log n)$ additional bits
by \cref{lem:cut-partition}.
This dominates the space for \cref{thm:dp-mwm} on
$H$, which has $O(\mu(G)/\delta)$ edges.
The positive crossing weights lie in $[\xi,W]$.
After scaling by $1/\xi$, \cref{lem:gw-mwm} uses
$O(\mu(G)\log(W/\delta)\log n)$ additional bits.
The total additional space is
$O(\mu(G)(\delta^{-2}+\log W)\log n)$ bits.

The construction of $H$ uses one stream pass to collect all internal edges,
and \cref{lem:gw-mwm} computes $J$ in one pass.
All other computations use stored data,
so the number of passes is $O(1)$.
\end{proof}

\subsubsection{Full algorithm and bounds}
\label{sec:streaming-full}
\Cref{alg:streaming-mwm} combines the two oracle implementations
with \cref{alg:phased-optimization,alg:fixed-phase}.

\begin{algorithm}[H]
\caption{Semi-streaming $(1-\eps)$-MWM}
\label{alg:streaming-mwm}
\small
\begin{algorithmic}[1]
\State $L\gets$ the weight of a $1/4$-MWM under $w$,
    using \cref{lem:gw-mwm}.
\State Initialize $x=0$ with empty support;
    set $t\gets1/(32\log W)$.
\While{$t\geq\eps/(2C\log W)$}
    \State Set $\delta,\eta$ as in \cref{sec:local-optimization}.
    \Loop
        \If{$\Eval(x,t,x)\geq3\delta L/4$}
            \State $d\gets-x$.
        \Else
            \State $d\gets\Step(x,t)$ using \cref{alg:streaming-step}.
            \If{$\Eval(x,t,d)>-\delta L/8$}
                \State \textbf{break}
            \EndIf
        \EndIf
        \State Update $x\gets x+\eta d$ and its support
            as in \cref{sec:streaming-eval}.
    \EndLoop
    \State $t\gets t/2$.
\EndWhile
\State $M\gets$ a $(1-\eps/2)$-MWM on $\supp(x)$
    under the original weights $w$, using \cref{thm:dp-mwm}.
\State \Return $M$.
\end{algorithmic}
\end{algorithm}

\begin{lemma}
\label{lem:streaming-base}
\Cref{alg:streaming-mwm} computes a $(1-\eps)$-MWM
using $O(\mu(G)\eps^{-2}\log W\log n)$ bits of space
and $O(\log W/\eps^2)$ passes.
\end{lemma}

\begin{proof}
The preliminary pass gives $\mu_w(G)/4\leq L\leq\mu_w(G)$.
By \cref{lem:streaming-eval,lem:streaming-step},
the analysis of \cref{lem:phased-optimization} at accuracy $\eps/2$
guarantees termination within $T$ iterations
with a fractional $(1-\eps/2)$-MWM.
Applying \cref{thm:dp-mwm} to its support returns a matching $M$ with
$w^\top\vone_M\geq(1-\eps/2)w^\top x
\geq(1-\eps/2)^2\mu_w(G)\geq(1-\eps)\mu_w(G)$.

Each iteration calls $\Step$ at most once,
using $O(1)$ passes by \cref{lem:streaming-step}.
The first-order evaluations, updates to $x$,
and final matching computation use stored data.
Including the preliminary pass for $L$,
the total number of passes is therefore $O(T)$.

The preliminary pass uses $O(\mu(G)\log W\log n)$ bits
by \cref{lem:gw-mwm}.
By \cref{lem:streaming-support,lem:streaming-eval,lem:streaming-step},
$O(\mu(G)(T+\delta^{-2}+\log W)\log n)$ bits suffice
for the optimization iterations.
Since $T=O(\eps^{-2}\log W)$
and every phase has $\delta\geq\eps/(2C)$,
the total space is $O(\mu(G)\eps^{-2}\log W\log n)$ bits.
The final call to \cref{thm:dp-mwm} uses
$O(|\supp(x)|\log n)$ bits of space
and fits within the same bound.
\end{proof}

\subsection{Removing the weight dependence}
\label{sec:streaming-weight-reduction}
We use the weight reduction of \cite{BernsteinCDLST25}
to reduce the weight range to $\eps^{-O(1)}$.

\begin{lemma}[{\cite[Theorem~6.11]{BernsteinCDLST25}}]
\label{lem:streaming-weight-reduction}
Suppose a semi-streaming algorithm computes a $(1-\eps)$-MWM
on an $n$-vertex, $m$-edge graph with weights in $[1,W]$
using $p(n,m,W,\eps)$ passes and $s(n,m,W,\eps)$ bits of space.
Then a $(1-O(\eps))$-MWM can be computed
using $p(n,m,\Theta(\eps^{-5}),\eps)$ passes and
$O(s(n,m,\Theta(\eps^{-5}),\eps)\cdot\lceil\log_{\eps^{-1}}W\rceil)$ bits of space.
The reduction preserves determinism.
\end{lemma}

\begin{proof}[Proof of \cref{thm:streaming}]
The MCM bound follows from \cref{lem:streaming-base} with $W=O(1)$.

For MWM, apply \cref{lem:streaming-weight-reduction}
to \cref{lem:streaming-base}
with accuracy a sufficiently small constant multiple of $\eps$,
using the latter directly when $W\leq\eps^{-5}$.
Each weight-class graph has maximum matching cardinality at most $\mu(G)$.
Since $W=n^{O(1)}$, this gives the second claim.
\end{proof}

\section{Incremental General Matching}
\label{sec:incremental}
In the incremental matching problem, we start with an empty graph on $n$ vertices
and maintain an explicit $(1-\eps)$-MWM after edge insertions.

We run \cref{alg:fixed-phase} with $t=\Theta(\eps/\log W)$
with a sufficiently small implied constant,
chosen so that $\delta=t\log W$ is a power of two,
and the step size $\eta$ from \cref{sec:local-optimization}.

The key observation in our incremental algorithm design is that
an edge insertion leaves the current fractional matching $x$ feasible,
so we can resume \cref{alg:fixed-phase} from $x$ after each insertion.
The cost of doing so depends on whether the algorithm changes $x$.
While $x$ is fixed, we need to maintain the oracles incrementally.
Otherwise, we take an optimization update and rebuild the cut partition and
the oracles. We first explain how to maintain the oracles when $x$
does not change.

\subsection{Oracles for a fixed iterate}
\label{sec:incremental-oracles}
For a fixed iterate $x$, we use its stored representation
and approximate duals from \cref{sec:streaming-eval},
together with the cut partition $\Q$ and weights $\widehat r$
from \cref{alg:streaming-step}.
The vertex duals and cut partition remain unchanged under insertions.

\paragraph{Step oracle.}
We maintain the matchings as follows,
ignoring edges with $\widehat r_e=0$.
\begin{enumerate}[label=\arabic*.,nosep]
    \item \emph{Internal matching.}
    Maintain a $(1-\delta)$-MWM within each part using \cref{thm:dp-mwm},
    and let $I$ be their union.
    An insertion affects at most one part,
    so recompute only that part's matching.
    \item \emph{Crossing matching.}
    Maintain a $1/6$-MWM $J$ using \cref{lem:incremental-mwm}.
    Initialize the instance on all current crossing edges,
    then insert each new crossing edge into it.
\end{enumerate}

\paragraph{First-order evaluations.}
We maintain the three scalar sums
$\widehat g^\top x$, $\widehat g^\top\vone_I$,
and $\widehat g^\top\vone_J$.
An insertion leaves $\widehat g^\top x$ unchanged,
since $x_e=0$ for the new edge.
Adjust the other two sums as $I$ and $J$ change.
These sums give each scalar first-order evaluation in $O(1)$ time.

\begin{lemma}
\label{lem:incremental-oracle}
For fixed $x$, oracles satisfying
\cref{def:first-order-oracle,def:step-oracle} with $c=O(1)$
can be maintained in $O(\delta^{-3}\log(2/\delta))$ time
per edge insertion.
\end{lemma}

\begin{proof}
The proofs of \cref{lem:streaming-eval,lem:streaming-step}
give the oracle guarantees,
with the crossing approximation factor $1/6$
changing only the absolute constant $c$.

Each part has fewer than $1/\delta$ vertices
by \cref{lem:cut-partition}.
Recomputing its matching therefore costs
$O(\delta^{-3}\log(2/\delta))$ by \cref{thm:dp-mwm}.
A crossing insertion costs $O(1)$ by \cref{lem:incremental-mwm},
and maintaining these sums fits within these bounds.
\end{proof}

\subsection{Full algorithm and bounds}
\label{sec:incremental-full}
After each edge insertion, we update the oracles as in
\cref{sec:incremental-oracles}
and resume \cref{alg:fixed-phase} from $x$.
When there is an optimization update, we rebuild the cut partition
and the oracles for the new iterate before continuing.

\begin{lemma}
\label{lem:incremental-progress}
There are $O(\log(nW)/(\eta\delta))$ optimization updates in total.
\end{lemma}

\begin{proof}
The per-update argument in the proof of \cref{lem:centering}
applies to every feasible iterate and gives a decrease
of $\Omega(\eta\delta\mu_w(G))$ in $F_t$ at each optimization update.
The arithmetic implementation of \cref{sec:streaming-eval}
preserves feasibility and a constant fraction of this decrease.

Insertions leave $F_t(x)$ unchanged, and
$0\leq F_t(0)-F_t(x)\leq w^\top x\leq\mu_w(G)$,
since $y_t(x)\geq y_t(0)$.
For any $k\geq0$, consider the optimization updates with
$2^k\leq\mu_w(G)<2^{k+1}$.
Each update decreases $F_t$ by $\Omega(\eta\delta2^k)$,
while their total decrease is at most $2^{k+1}$.
There are therefore $O(1/(\eta\delta))$ such updates.
Since $1\leq\mu_w(G)\leq nW$ whenever an update occurs,
there are $O(\log(nW))$ possible ranges, proving the bound.
\end{proof}

Now we give the full theorem.

\begin{samepage}
\incrementaltheorem*
\end{samepage}

\begin{proof}
We prove the approximation guarantee by induction over insertions.
Assume the previous output $M$ is a $(1-\eps)$-MWM before inserting $e$.
For the estimate required by \cref{alg:fixed-phase},
use $L=\max\{w(M),w_e\}$.
Removing $e$ from any matching in $G$
leaves a matching in the previous graph, so
$L\leq\mu_w(G)\leq w(M)/(1-\eps)+w_e<4L$.
The stopping argument in the proof of \cref{lem:centering}
gives $F_t(x)-F_t^*\leq2c^2\delta\mu_w(G)$.
Our choice of $t$ and \cref{lem:potential-bias}
then imply that $x$ is a fractional $(1-\eps/4)$-MWM.
Its support therefore contains an integral matching
of weight at least $w^\top x$.
We retain the previous output $M$
if $w(M)\geq(1-\eps/2)w^\top x$.
Otherwise, we recompute $M$ as a $(1-\eps/4)$-MWM
on $\supp(x)$ using \cref{thm:dp-mwm}.
Whether we retain $M$ or recompute it,
$w(M)\geq(1-\eps/2)w^\top x
\geq(1-\eps/2)(1-\eps/4)\mu_w(G)
\geq(1-\eps)\mu_w(G)$.
This completes the induction and justifies the estimates $L$.

Initialization takes $O(n)$ time.
Processing the $m$ insertions for the oracle takes
$O(m\delta^{-3}\log(2/\delta))$ time
by \cref{lem:incremental-oracle}.
The recomputation rule ensures that each new matching after the first
has weight greater than $(1-\eps/4)/(1-\eps/2)=1+\Omega(\eps)$
times the previous output weight.
Since these weights lie between $1$ and $nW$,
we compute a new matching $O(\eps^{-1}\log(nW))$ times.
By \cref{thm:dp-mwm}, these computations take
$O(m\eps^{-2}\log(2/\eps)\log(nW))$ time in total.

By \cref{lem:incremental-progress}, there are
$O(\log(nW)/(\eta\delta))$ optimization updates.
After each such update, we scan $G$ to collect internal edges
and reinitialize the crossing instance.
Reconstructing the cut partition and computing all internal matchings take
$O(|E|+n\delta^{-3})$ time by
\cref{lem:cut-partition,lem:internal-edge-bound,thm:dp-mwm}.
The duals and evaluations take $O(|E|+n\log(W/\delta))$ time
by \cref{lem:streaming-eval}.
Including maintenance of the stored iterate,
each optimization update therefore costs $O(m+n\delta^{-3}+n\log W)$ time.
Substituting $\delta=\Theta(\eps)$ and $\eta=\Theta(\eps^2/\log W)$
gives the claimed total time $O((m\eps^{-3}+n\eps^{-6})\log^3 n)$.
\end{proof}

\section{Fully Dynamic Low-Loss Reduction to Bipartite MCM}
\label{sec:dynamic}
In the fully dynamic matching problem,
we maintain an explicit $(1-\eps)$-MWM in a graph
undergoing edge insertions and deletions.
In this section, we give the following
low-loss reduction from MWM in general graphs
to MCM in bipartite graphs.\footnote{In
\cref{thm:dynamic-reduction}, we take $U_b(n,m,\rho)$
to be nondecreasing in $n,m$
and nonincreasing in $\rho$, without loss of generality.
We also use the bound
$U_b=\Omega(1/\rho)$, which follows by repeatedly
deleting both end edges of a path of odd length $\Theta(1/\rho)$
and then reinserting them.
With a sufficiently small implied constant,
a $(1-\rho)$-MCM must be maximum,
so these updates force it to switch between the odd- and even-indexed edges;
see the proof of \cite[Fact~2.3]{BhattacharyaKSW24}.}

\begin{samepage}
\dynamicreductiontheorem*
\end{samepage}

We first present an algorithm
with $\otilde(\eps^{-6})$ overhead.
Using the standard periodic rebuilding framework,
we reduce the task to repeatedly computing a $(1-\eps/2)$-MWM
using dynamic bipartite MCM queries.
We adapt the oracle implementations of
\cref{sec:streaming-eval,sec:streaming-step}
to compute each matching without scanning the whole graph.
A particular challenge is to compute the crossing matching
as the approximate weights $-\widehat g_{uv}=w_{uv}-\widehat y_u-\widehat y_v$
change with updates to the iterate.
To handle the changes,
we combine graph unfolding in \cite{KaoLST01,BernsteinDL21}
and the vertex toggle idea in \cite{BernsteinC26}.
We then apply the weight reduction of \cite{Chen26}
to remove the dependence on $W$ from the overhead.

We make two further improvements.
First, we reuse the cut partition from \cref{sec:cut-partition}
across iterations to improve the overhead
to $\otilde(\eps^{-5})$.
Second, we customize the matching composition of \cite{Chen26}
to combine matchings in different weight windows
computed at different update times,
allowing less frequent recomputation and reducing the overhead
to $\otilde(\eps^{-4})$.

\subsection{Periodic rebuilding}
\label{sec:dynamic-baseline}
Let $G_0,G_1,\ldots$ denote the sequence of graphs,
where $G_0$ is empty and $G_i$ is obtained from $G_{i-1}$
by one edge insertion or deletion.
The following standard bound shows how long a matching remains
approximately optimal between rebuilds.

\begin{lemma}
\label{lem:dynamic-periodic-rebuilding}
Let $M$ be a $(1-\eps/2)$-MWM of $G_i$.
For every $j$ with $0\leq j-i\leq\eps\mu_w(G_i)/(8W)$,
$M\cap G_j$ is a $(1-\eps)$-MWM of $G_j$.
\end{lemma}

\begin{proof}
The matching $M\cap G_j$ has weight at least
$w(M)-(j-i)W$, while $\mu_w(G_j)\leq\mu_w(G_i)+(j-i)W$.
Thus $\mu_w(G_j)-w(M\cap G_j)
    \leq \eps\mu_w(G_i)/2+2(j-i)W
    \leq 3\eps\mu_w(G_i)/4$.
On the other hand,
$\mu_w(G_j)\geq\mu_w(G_i)-(j-i)W
\geq(1-\eps/8)\mu_w(G_i)
\geq3\mu_w(G_i)/4$.
Therefore $w(M\cap G_j)\geq(1-\eps)\mu_w(G_j)$.
\end{proof}

At each rebuild, we compute a fresh $(1-\eps/2)$-MWM.
We bound the recomputation cost in terms of $\mu_w(G)$ and $\mu(G)$
to amortize it over the interval in \cref{lem:dynamic-periodic-rebuilding}.
Within each recomputation, we maintain $x$ and its support
as in \cref{sec:streaming-eval}.
The support bound of \cref{lem:streaming-support} therefore applies.
We also use the cut construction and internal matching
from \cref{alg:streaming-step}.
The dynamic implementation recovers internal edges through adjacency queries
and uses the given dynamic bipartite MCM algorithm for crossing matchings.

\subsection{Dynamic crossing matching with vertex dual updates}
\label{sec:dynamic-auxiliary}
We compute the crossing matching $J$
using the approximate duals from \cref{sec:streaming-eval},
with weights $-\widehat g_{uv}=w_{uv}-\widehat y_u-\widehat y_v$.
Input edge updates and changes to $\Q$ change the crossing edge set.
Updates to $x$ change the vertex duals
and hence the weights of incident edges.
We first give a data structure for integer edge weights and vertex duals,
then round the edge weights and duals to apply it.

\begin{problem}
\label{prob:dynamic-dual-mwm}
Fix an integer bound $W\geq1$.
Consider a fully dynamic graph $G=(V,E,w)$
with integer edge weights $w_e\in\{0,\ldots,W\}$
and integer vertex duals $y_v\in\{0,\ldots,W\}$.
Initially $G$ is empty and $y=0$.
Design a data structure that supports the following operations:
\begin{enumerate}[label=\arabic*.,nosep]
    \item \emph{Edge update.} Insert or delete an edge $uv$ of weight $w_{uv}$.
    \item \emph{Dual update.} Change $y_v$ to an integer
        $y'_v\in\{0,\ldots,W\}$.
    \item \emph{Matching query.} Return an approximate MWM in $G$
        under weights $k_{uv}=w_{uv}-y_u-y_v$.
\end{enumerate}
\end{problem}

\subsubsection{Graph unfolding and vertex toggles}
\label{sec:dynamic-unfolding-static}
To solve \cref{prob:dynamic-dual-mwm},
we reduce approximate MWM under weights $k_{uv}=w_{uv}-y_u-y_v$
to approximate MCM in a bipartite unweighted graph,
using graph unfolding \cite{KaoLST01,BernsteinDL21}
and vertex toggles \cite{BernsteinC26}.

\begin{definition}
\label{def:dynamic-auxiliary-graph}
Fix an integer bound $W$.
For a graph $G=(V,E,w)$ with integer edge weights $w_e\in\{0,\ldots,W\}$,
the \emph{unfolded graph} $\phi(G)=(V^\prime,E^\prime)$ is bipartite and unweighted,
with $V^\prime\defeq\{v_i^L,v_i^R:v\in V,\ 1\leq i\leq W\}$
and $
    E^\prime\defeq\{u_i^Lv_{w_{uv}+1-i}^R,\ u_i^Rv_{w_{uv}+1-i}^L:
        uv\in E,\ 1\leq i\leq w_{uv}\}$.
For a subgraph $F\subseteq\phi(G)$,
its \emph{refolding} $\mathcal R(F)$ consists of the original edges in $G$
corresponding to $F$.
\end{definition}

For integer vertex duals $y_v\in\{0,\ldots,W\}$,
define the \emph{toggle set} $S(y)=\{v_i^L,v_i^R:v\in V,\ 1\leq i\leq y_v\}$.
We show that the refolding of a large unweighted
matching in $\phi(G)[V^\prime\setminus S(y)]$ contains
a large weighted matching in $G$ under the weights $k_{uv}=w_{uv}-y_u-y_v$.

\begin{lemma}
\label{lem:dynamic-refolding}
For every $(\alpha,\beta)$-MCM $M$ of $\phi(G)[V^\prime\setminus S(y)]$,
\[
    \mu_k(\mathcal R(M))
    \geq \frac{\alpha}{2}\mu_k(G)-\frac{\beta}{4}.
\]
\end{lemma}

\begin{proof}
Each edge $uv$ of an MWM under $k$
contributes $k_{uv}$ disjoint edge copies in each orientation
with endpoints outside $S(y)$.
Thus $\mu(\phi(G)[V^\prime\setminus S(y)])\geq2\mu_k(G)$,
and $|M|\geq2\alpha\mu_k(G)-\beta$.

It remains to show $|M|\leq4\mu_k(\mathcal R(M))$.
Fix an MWM in $\mathcal R(M)$ under $k$.
For each vertex $v$, let $z_v$ be the weight of its incident edge in this matching,
or zero if $v$ is unmatched.
We have $z_u+z_v\geq k_{uv}$ for every $uv\in E(\mathcal R(M))$;
otherwise, replacing the matching edges incident to $u,v$ by $uv$
would increase its weight.
For an edge copy $u_i^Lv_j^R$ in $M$,
we have $i>y_u$, $j>y_v$, and
$(i-y_u)+(j-y_v)=k_{uv}+1$.
Hence either $i-y_u\leq z_u$ or $j-y_v\leq z_v$.
The same holds for the other orientation.
Thus the copies $v_i^L,v_i^R$ with $y_v<i\leq y_v+z_v$
form a vertex cover of $M$ of size at most
$2\sum_v z_v=4\mu_k(\mathcal R(M))$.
Since $M$ is a matching, $|M|\leq4\mu_k(\mathcal R(M))$.
\end{proof}

We then use the construction of \cite{BernsteinC26}
to handle vertex toggles using auxiliary edges.

\begin{lemma}
\label{lem:dynamic-vertex-toggles}
For each $u\in V^\prime$, add a new vertex $\bar u$.
Consider the auxiliary graph with vertex set
$V^\prime\cup\{\bar u:u\in V^\prime\}$
and edge set $E^\prime\cup\{u\bar u:u\in S(y)\}$.
For any $(1-\rho)$-MCM $M$ in this auxiliary graph,
the restriction $M\cap E(\phi(G)[V^\prime\setminus S(y)])$
is a $(1-\rho,\rho|S(y)|)$-MCM of $\phi(G)[V^\prime\setminus S(y)]$.
\end{lemma}

\begin{proof}
An MCM of $\phi(G)[V^\prime\setminus S(y)]$
extends to a matching in the auxiliary graph
by adding all $|S(y)|$ auxiliary edges.
Every edge of $M$ outside $\phi(G)[V^\prime\setminus S(y)]$
meets $S(y)$, so restricting $M$ discards at most $|S(y)|$ edges.
Since $M$ is a $(1-\rho)$-MCM in the auxiliary graph, we have
$|M\cap E(\phi(G)[V^\prime\setminus S(y)])|
        \geq |M|-|S(y)|
        \geq (1-\rho)\mu(\phi(G)[V^\prime\setminus S(y)])-\rho|S(y)|$.
\end{proof}

Based on the above constructions,
we use a dynamic MCM algorithm for bipartite graphs to implement
a data structure that solves \cref{prob:dynamic-dual-mwm}.

\begin{lemma}
\label{lem:dynamic-unfolding}
Fix $0<\rho\leq1/4$. Suppose that
there is a fully dynamic algorithm that,
on a bipartite graph with at most $n$ vertices and $m$ edges,
explicitly maintains a $(1-\rho)$-MCM against adaptive adversaries
in amortized update time $U_b(n,m,\rho)$.

Then there is a data structure for \cref{prob:dynamic-dual-mwm},
on a graph with at most $n$ vertices and $m$ edges, with
amortized edge update time for an edge $e$
\[
    O(w_e)\cdot U_b(O(nW),O((m+n)W),\rho),
\]
amortized dual update time
\[
    O(|y'_v-y_v|)\cdot U_b(O(nW),O((m+n)W),\rho).
\]
Each matching query
returns a $(1/3,O(\rho\|y\|_1))$-MWM
under weights $k$ in $O(\mu_w(G)+\|y\|_1)$ time.
The returned matching only consists of edges with $k_e>0$.
\end{lemma}

\begin{proof}
Maintain the auxiliary graph of \cref{lem:dynamic-vertex-toggles}
using the given bipartite MCM algorithm at accuracy $\rho$.
This graph has $4nW$ vertices and $O((m+n)W)$ edges.
Inserting or deleting an original edge $e$
inserts or deletes its $2w_e$ unfolded copies.
A dual update $y_v\gets y'_v$ inserts or deletes the auxiliary edges
incident to $v_i^L,v_i^R$,
which uses $2|y'_v-y_v|$ edge updates and proves the update bounds.

For a matching query, restrict the matching maintained
in the auxiliary graph
to $\phi(G)[V^\prime\setminus S(y)]$
to obtain a matching $M$ in the unfolded graph.
Each edge copy in $M$ corresponds to an original edge in $G$,
but these original edges may share endpoints.
Initialize $M^\prime=\varnothing$
and process the edge copies in $M$ in decreasing order of $k_e$,
where $e$ is the corresponding original edge.
Add $e$ to $M^\prime$ whenever both its endpoints are unmatched in $M^\prime$.
Return the resulting matching $M^\prime\subseteq\mathcal R(M)$ in the original graph $G$.

For every edge copy of $uv$ in $M$, its endpoint indices satisfy
$i\geq y_u+1$, $j\geq y_v+1$, and $i+j=w_{uv}+1$.
Thus $k_{uv}=i+j-1-y_u-y_v\geq1$.

By \cref{lem:dynamic-vertex-toggles} and $|S(y)|=2\|y\|_1$,
we have $|M|\geq(1-\rho)\mu(\phi(G)[V^\prime\setminus S(y)])-2\rho\|y\|_1$.
Assign each vertex the weight $k_e$ of its incident edge $e\in M^\prime$,
or zero if it is unmatched.
These labels sum to at least $k_{uv}$ on each refolded edge $uv$,
since every edge rejected by the greedy algorithm
meets a previously selected edge of at least its weight.
The vertex-cover argument in the proof of \cref{lem:dynamic-refolding}
therefore gives $|M|\leq4k(M^\prime)$.
The proof of \cref{lem:dynamic-refolding} also shows
$\mu(\phi(G)[V^\prime\setminus S(y)])\geq2\mu_k(G)$.
Combining these inequalities gives
$k(M^\prime)\geq\frac{|M|}{4}
    \geq\frac{1-\rho}{2}\mu_k(G)-\frac{\rho}{2}\|y\|_1$.

Obtaining $M$ takes $O(|M|+\|y\|_1)$ time.
We have $|M|\leq4k(M^\prime)\leq4\mu_w(G)$.
The positive integer weights $k_e$ are at most $\mu_w(G)$,
so bucket sorting and greedy selection take $O(|M|+\mu_w(G))$ time.
The total query time is thus $O(\mu_w(G)+\|y\|_1)$.
\end{proof}

\subsubsection{Computing the crossing matching}
\label{sec:dynamic-crossing}
We now compute the crossing matching
w.r.t.\ $\widehat y$ as in \cref{lem:streaming-eval}.
To apply \cref{lem:dynamic-unfolding} for integer weights and duals,
we define
$w^\Delta_e=\max\{\lfloor w_e/\Delta\rfloor-2,0\}$,
and $\widehat y^\Delta_{t,v}
=\min\{\lfloor\widehat y_v/\Delta\rfloor,\lceil W/\Delta\rceil\}$
for $\Delta>0$.
The following lemma transfers its guarantee to the true weights $-\grad F_t(x)$.

\begin{lemma}
\label{lem:dynamic-crossing-rounding}
For $\Delta>\delta/128$, the rounded weights and duals satisfy:
\begin{enumerate}[label=\arabic*.,nosep]
    \item If $\grad F_t(x)^\top x\leq\delta\mu_w(G)$,
        then $\Delta\|\widehat y^\Delta_t\|_1\leq3\mu_w(G)$.
    \item Let $F\subseteq G$, and let $M$ be an $(\alpha,\beta)$-MWM in $F$
        under weights $w^\Delta_{uv}-\widehat y^\Delta_{t,u}-\widehat y^\Delta_{t,v}$,
        consisting of positive-weight edges.
        Then $M$ is an $(\alpha,4\alpha\Delta\mu(G)+\Delta\beta)$-MWM
        in $F$ under weights $-\grad F_t(x)$,
        and $-\grad F_t(x)_e>0$ for every $e\in M$.
\end{enumerate}
\end{lemma}

\begin{proof}
For the first claim, $x_v\leq1/2$ implies
$\widehat y_v<\delta/128<\Delta$.
Thus $\widehat y^\Delta_{t,v}>0$ only if $x_v>1/2$.
There are at most $4\mu(G)$ such vertices,
so the same dual approximation bound gives
\[
    \Delta\|\widehat y^\Delta_t\|_1
    \leq\sum_{v:x_v>1/2}\widehat y_v
    \leq2\sum_vx_vy_{t,v}(x)+\delta\mu(G)
    =2(\grad F_t(x)^\top x+w^\top x)+\delta\mu(G)
    \leq3\mu_w(G),
\]
where the last inequality uses $w^\top x\leq\mu_w(G)$,
$\mu(G)\leq\mu_w(G)$, and $\delta\leq1/32$.

For the second claim, we first bound the loss from rounding on any edge $uv$:
$\Delta(w^\Delta_{uv}-\widehat y^\Delta_{t,u}-\widehat y^\Delta_{t,v})
    \geq-\widehat g_{uv}-3\Delta
    \geq-\grad F_t(x)_{uv}-4\Delta$,
where the second inequality uses \cref{lem:streaming-eval}
and $\Delta>\delta/128$.

To lower-bound the true weight of $M$,
we use the positivity of its rounded edge weights.
For each $uv\in M$, both endpoint duals are less than
$w^\Delta_{uv}\leq\lceil W/\Delta\rceil$,
so neither is capped.
Moreover, $w^\Delta_{uv}>0$ implies $\Delta w^\Delta_{uv}\leq w_{uv}-2\Delta$.
Since $\Delta\widehat y^\Delta_{t,v}>\widehat y_v-\Delta$ at both endpoints,
$0<\Delta(w^\Delta_{uv}-\widehat y^\Delta_{t,u}-\widehat y^\Delta_{t,v})
    \leq-\widehat g_{uv}\leq-\grad F_t(x)_{uv}$.

Let $M^*$ be an MWM in $F$ under weights $-\grad F_t(x)$.
Combining these bounds with the $(\alpha,\beta)$-guarantee of $M$ gives
\[
\begin{aligned}
    -\grad F_t(x)^\top\vone_M
    &\geq\Delta\sum_{uv\in M}
        (w^\Delta_{uv}-\widehat y^\Delta_{t,u}-\widehat y^\Delta_{t,v})\\
    &\geq\alpha\Delta\sum_{uv\in M^*}
        (w^\Delta_{uv}-\widehat y^\Delta_{t,u}-\widehat y^\Delta_{t,v})-\Delta\beta
    \geq-\alpha\grad F_t(x)^\top\vone_{M^*}
        -4\alpha\Delta\mu(G)-\Delta\beta,
\end{aligned}
\]
where the last inequality uses $|M^*|\leq\mu(G)$.
This proves the second claim.
\end{proof}

To compute the crossing matching, we choose $\Delta$ so that
rounding incurs $O(\delta\mu_w(G))$ additive error
and each matching query takes $O(\mu(G)/\delta)$ time.

\begin{lemma}
\label{lem:dynamic-crossing}
Let $x\in\P_{G,\delta}$ satisfy
$\grad F_t(x)^\top x\leq\delta\mu_w(G)$,
and let $\widehat y$ satisfy the dual approximation bound of \cref{lem:streaming-eval}.
Let $H\subseteq G$ and $0<\rho\leq\delta$.
Choose $\Delta=\Theta\!\left(\frac{\delta\mu_w(G)}{\mu(G)}\right)$
so that $\delta/128<\Delta\leq W$.
A data structure of \cref{lem:dynamic-unfolding}
representing $G\setminus H$ with edge weights $w^\Delta$,
vertex duals $\widehat y^\Delta_t$,
integer bound $\lceil W/\Delta\rceil$, and accuracy $\rho$
returns a $(1/3,O(\delta\mu_w(G)))$-MWM in $G\setminus H$
under weights $-\grad F_t(x)$ in $O(\mu(G)/\delta)$ query time.
The returned matching contains only edges with $-\grad F_t(x)_e>0$.
\end{lemma}

\begin{proof}
The choice of $\Delta$ gives
$\Delta\mu(G)=O(\delta\mu_w(G))$
and $\mu_w(G)/\Delta=O(\mu(G)/\delta)$.
By \cref{lem:dynamic-unfolding}, the query returns a
$(1/3,O(\rho\|\widehat y^\Delta_t\|_1))$-MWM
under weights $w^\Delta_{uv}-\widehat y^\Delta_{t,u}-\widehat y^\Delta_{t,v}$,
containing only positive-weight edges.
\Cref{lem:dynamic-crossing-rounding} bounds
$\|\widehat y^\Delta_t\|_1=O(\mu_w(G)/\Delta)$
and gives the claimed approximation and positivity under $-\grad F_t(x)$.
Since $\mu_{w^\Delta}(G\setminus H)\leq\mu_w(G)/\Delta$,
\cref{lem:dynamic-unfolding} gives the query time.
\end{proof}

\subsection{Dynamic step oracle}
\label{sec:dynamic-step}
We use the internal and crossing directions of \cref{alg:streaming-step}.
We recover the internal graph $H$ through adjacency queries
and compute its matching $I$ as in that algorithm.
The data structure in \cref{lem:dynamic-unfolding} computes the crossing matching $J$.
We store the input graph to support constant-time
adjacency and edge-weight queries.
\Cref{alg:dynamic-step} combines these computations
with the approximate duals and $\Eval$ from \cref{sec:streaming-eval}.

Let $\delta=t\log W$, $0<\rho\leq\delta$,
and $\Delta=\Theta(\delta\mu_w(G)/\mu(G))$
so that $\delta/128<\Delta\leq W$,
as in \cref{lem:dynamic-crossing}.
We assume access to a data structure $\D^\Delta$
in \cref{lem:dynamic-unfolding}
with integer bound $\lceil W/\Delta\rceil$ and accuracy $\rho$.
After extracting the crossing matching,
we reset the vertex duals to zero
but leave the internal edges excluded
so that the accelerated algorithm in \cref{subsec:dynamic-cut-reuse}
can reuse the crossing graph
across oracle calls.

\begin{algorithm}[H]
\caption{Dynamic $\Step(x,t)$}
\label{alg:dynamic-step}
\small
\begin{algorithmic}[1]
\Require $x\in\P_{G,\delta}$ with
    $\grad F_t(x)^\top x\leq\delta\mu_w(G)$ and
    the current internal graph $H_{\mathrm{old}}$.
\Require $\D^\Delta$ representing $G\setminus H_{\mathrm{old}}$
    with weights $w^\Delta$ and zero vertex duals.
\State Compute $\widehat y$ as in \cref{lem:streaming-eval}
    and its rounded duals $\widehat y^\Delta_t$ at the endpoints of $\supp(x)$.
\State Apply \cref{lem:cut-partition} to $x$ to obtain $\Q$.
\State Compute $H=(V,\bigcup_{Q\in\Q}E[Q])$ through adjacency queries within each part.
\State $\D^\Delta.\mathtt{EdgeUpdate}(\mathtt{insert},e,w^\Delta_e)$
    for each $e\in E(H_{\mathrm{old}})$.
\State $\D^\Delta.\mathtt{EdgeUpdate}(\mathtt{delete},e)$
    for each $e\in E(H)$.
\State $H_{\mathrm{old}}\gets H$.
\State $I\gets$ a $(1-\delta)$-MWM in $H$
    under weights $\widehat r$ from \cref{alg:streaming-step},
    discarding zero edges and using \cref{thm:dp-mwm}.
\State $\D^\Delta.\mathtt{DualUpdate}(v,\widehat y^\Delta_{t,v})$
    for each vertex $v$ incident to $\supp(x)$.
\State $J\gets\D^\Delta.\mathtt{MatchingQuery}()$.
\State $\D^\Delta.\mathtt{DualUpdate}(v,0)$
    for each vertex $v$ incident to $\supp(x)$.
\State \Return $\argmin_{d\in\{(1-2\delta)\vone_I-x,\vone_J\}}\Eval(x,t,d)$.
\end{algorithmic}
\end{algorithm}

\begin{lemma}
\label{lem:dynamic-step}
Suppose there is a fully dynamic algorithm that,
on a bipartite graph with at most $n$ vertices and $m$ edges,
explicitly maintains a $(1-\rho)$-MCM against adaptive adversaries
in amortized update time $U_b(n,m,\rho)$.
Then \cref{alg:dynamic-step}
returns a direction satisfying \cref{def:step-oracle}
in amortized time
\[O\Bigl(|\supp(x)|+\mu(G)(\delta^{-3}+\log(W/\delta))
        +\mu(G)\delta^{-2}\cdot
        U_b\left(O(nW\Delta^{-1}),O((m+n)W\Delta^{-1}),\rho\right)\Bigr).
\]
\end{lemma}

\begin{proof}
At vertices not incident to $\supp(x)$,
$\widehat y_v=\delta/8192<\Delta$, so $\widehat y^\Delta_{t,v}=0$.
The stored duals are initially zero,
so the edge and dual updates make $\D^\Delta$ represent $G\setminus H$
with the current rounded duals at the matching query.
By \cref{lem:dynamic-crossing}, $J$ is a
$(1/3,O(\delta\mu_w(G)))$-MWM under weights $-\grad F_t(x)$
and contains only positive-weight edges.
Applying \cref{lem:cut-partition} to $x$
guarantees odd-set feasibility of $x+\eta\vone_J$.
By the proof of \cref{lem:streaming-step},
$I$ is a $(1-\delta,\delta\mu_w(G))$-MWM in $H$ under weights $-\grad F_t(x)$.
The proof of \cref{lem:matching-directions} absorbs this error
and the $\delta\mu_w(G)/512$ error in selecting the direction with $\Eval$,
giving the claimed guarantee.
The final dual updates restore zero duals.

For the choices $\Delta\mu(G)\leq\delta\mu_w(G)/4$ and $\rho\leq\delta/2$
used below, $c=4$ suffices.
Indeed, the proofs of \cref{lem:dynamic-unfolding,lem:dynamic-crossing-rounding}
bound the crossing error by
$\frac43\Delta\mu(G)+\frac32\rho\mu_w(G)\leq2\delta\mu_w(G)$.
Together with the internal matching and direction-selection errors above,
the proof of \cref{lem:matching-gain-decomposition} gives $c=4$.

\Cref{lem:cut-partition} constructs $\Q$ in
$O(|\supp(x)|+\mu(G)\delta^{-3})$ time.
The proof of \cref{lem:internal-edge-bound} shows that
its nonsingleton parts have total size $O(\mu(G))$
and each has fewer than $1/\delta$ vertices.
Thus recovering $H$ takes $O(\mu(G)\delta^{-1})$ adjacency queries.
Computing the duals and evaluating the two directions takes
$O(|\supp(x)|+\mu(G)\log(W/\delta))$ time
by \cref{lem:streaming-eval}.
Since $|E(H)|=O(\mu(G)\delta^{-1})$,
evaluating $\widehat r$ on $E(H)$ and computing $I$ takes
$O(\mu(G)\delta^{-2}\log(1/\delta))$ time by \cref{thm:dp-mwm}.

We next bound the time for restoring $E(H_{\mathrm{old}})$
and deleting $E(H)$ in $\D^\Delta$.
By the edge-update guarantee of \cref{lem:dynamic-unfolding},
it suffices to bound the total rounded weight of these updates.
Both internal graphs have maximum degree less than $1/\delta$.
In either graph, consider a greedy matching obtained
by processing edges in decreasing order of original weight.
Every edge meets a selected edge of at least its weight,
and each selected edge meets at most $2/\delta$ edges.
Since the selected matching has weight at most $\mu_w(G)$,
each internal graph has total edge weight $O(\mu_w(G)\delta^{-1})$.

Since $w^\Delta_e\leq w_e/\Delta$,
the sum of rounded edge weights over all insertions and deletions is
$O(\mu_w(G)/(\delta\Delta))=O(\mu(G)\delta^{-2})$.
These operations require $O(\mu(G)\delta^{-2})$ auxiliary edge updates
and $O(\mu(G)\delta^{-1})$ time to process the original edge lists.

By \cref{lem:dynamic-crossing-rounding},
$\|\widehat y^\Delta_t\|_1=O(\mu_w(G)/\Delta)=O(\mu(G)\delta^{-1})$.
Setting and resetting the duals therefore requires
$O(\mu(G)\delta^{-1})$ auxiliary edge updates,
and the crossing query takes $O(\mu(G)\delta^{-1})$ time
by \cref{lem:dynamic-crossing}.
In total, the call uses $O(\mu(G)\delta^{-2})$ auxiliary edge updates
and $O(|\supp(x)|+\mu(G)(\delta^{-3}+\log(W/\delta)))$ other operations.
The auxiliary graph has $O(nW\Delta^{-1})$ vertices
and $O((m+n)W\Delta^{-1})$ edges by \cref{lem:dynamic-unfolding}.
Applying the amortized update bound $U_b$ proves the stated running time.
\end{proof}

\subsection{Full weight-dependent algorithm}
\label{sec:dynamic-completion}
\Cref{alg:dynamic-mwm} uses the phased optimization
and final matching computation of \cref{alg:streaming-mwm}
with the dynamic oracles above.
Maintain instances $\D^{\Delta_\ell}$ of \cref{lem:dynamic-unfolding}
for values $\Delta_\ell$ forming a geometric sequence with ratio two,
starting at a sufficiently small constant multiple of $\eps$
and extending up to $W$,
with rounded weights $w^{\Delta_\ell}$ and accuracy $\rho=\Theta(\eps)$.
Send each input edge update to every instance with its rounded weight,
so between periodic rebuilds,
all instances represent the full graph.

For selection of $\Delta$, we obtain constant-approximate estimates
$L$ of $\mu_w(G)$ and $r$ of $\mu(G)$.
A zero-dual query with a sufficiently small constant $\Delta$
returns a matching whose weight under $w$ gives $L$.
A unit-weight instance $\D^{\mathrm{unit}}$ gives $r$
by the same guarantee of \cref{lem:dynamic-unfolding}.
In a phase with $\delta=t\log W$,
we choose $\Delta\in\{\Delta_\ell\}$ satisfying $\Delta=\Theta(\delta L/r)$.

\begin{algorithm}[H]
\caption{Fully dynamic $(1-\eps)$-MWM using dynamic MCM queries}
\label{alg:dynamic-mwm}
\small
\begin{algorithmic}[1]
\Require $G=(V,E,w)$, $0<\eps\leq1/2$,
$\mu_w(G)/4\leq L\leq\mu_w(G)$, $\mu(G)\leq r\leq3\mu(G)$.
\Require Maintained instances $\D^{\Delta_\ell}$ on $G$ with zero duals.
\State If $E=\varnothing$, return $\varnothing$.
\State Initialize $x=0$ with empty support.
    Set $t\gets1/(32\log W)$.
\While{$t\geq \eps/(2C\log W)$}
    \State Set $\delta,\eta$ as in \cref{sec:local-optimization}.
        Choose $\Delta\in\{\Delta_\ell\}\cap(\delta L/(8r),\delta L/(4r)]$.
    \State $H_{\mathrm{old}}\gets(V,\varnothing)$.
    \State $x\gets$ \cref{alg:fixed-phase}$(x,t,L)$
        using \cref{alg:dynamic-step} with $\D^\Delta,H_{\mathrm{old}}$,
        maintained as in \cref{sec:streaming-eval}.
    \State $\D^\Delta.\mathtt{EdgeUpdate}(\mathtt{insert},e,w^\Delta_e)$
        for each $e\in E(H_{\mathrm{old}})$.
    \State $t\gets t/2$.
\EndWhile
\State $M\gets$ a $(1-\eps/2)$-MWM on $\supp(x)$
    under the original weights $w$,
    using \cref{thm:dp-mwm}.
\State \Return $M$.
\end{algorithmic}
\end{algorithm}

\begin{lemma}
\label{lem:dynamic-periodic-bound}
Given a bipartite unweighted matching algorithm
with update bound $U_b(n,m,\rho)$ as above, for every $0<\eps\leq1/4$,
there is a fully dynamic algorithm that,
on an initially empty general graph with at most $n$ vertices and $m$ edges
and weights in $[1,W]$,
explicitly maintains a $(1-\eps)$-MWM in amortized update time
\[
    O\!\left(W\eps^{-5}\log^2 W\cdot
        U_b\!\left(O(nW\eps^{-1}),O((m+n)W\eps^{-1}),\Theta(\eps)\right)\right).
\]
The algorithm is deterministic when the bipartite algorithm is deterministic.
\end{lemma}

\begin{proof}
For \cref{alg:dynamic-mwm}, set $C=256$, $\rho=\eps/(4C)$, and
$\Delta_\ell=2^\ell\eps/(1024C)\leq W$ for $\ell=0,1,\ldots$.
Maintain the instances described above with integer bounds
$\lceil W/\Delta_\ell\rceil$ and $1$
for $\D^{\Delta_\ell}$ and $\D^{\mathrm{unit}}$, respectively.

A zero-dual query of a maintained instance with $1/32<\Delta\leq1/16$
returns a matching of original weight
$L\geq(\mu_w(G)-3\Delta\mu(G))/3\geq\mu_w(G)/4$
by \cref{lem:dynamic-unfolding}.
The unit-weight query returns a $1/3$-MCM $A$,
so $r=3|A|$ satisfies $\mu(G)\leq r\leq3\mu(G)$.
Both estimates take $O(\mu_w(G))$ time.

The estimate bounds give
$\Delta>\delta\mu_w(G)/(96\mu(G))\geq\delta/96$
and $\Delta\leq\delta W/4$.
Since $\delta\geq\eps/(2C)$, the selection interval lies within $(\Delta_0,W]$.
Its endpoint ratio is two, so it contains a value $\Delta_\ell$
satisfying \cref{lem:dynamic-crossing}.
Since $\delta\geq\eps/(2C)$, we also have $\rho\leq\delta/2$.
The shrinking test and \cref{def:first-order-oracle} give
$\grad F_t(x)^\top x<\delta L\leq\delta\mu_w(G)$,
so \cref{lem:dynamic-step} applies at every oracle call.
The error bounds in the proof of \cref{lem:dynamic-step}
give $c=4$, for which $C=256$ suffices.
The convergence argument and final matching computation
in \cref{lem:streaming-base} then give a $(1-\eps)$-MWM.
Dual resets and phase-end edge restorations return every instance to $G$
with zero duals.

Summing \cref{lem:dynamic-unfolding} over the instances $\D^{\Delta_\ell}$
bounds the amortized maintenance cost per input edge update by
\[
    O(W\eps^{-1})\cdot
    U_b\left(O(nW\eps^{-1}),O((m+n)W\eps^{-1}),\eps/(4C)\right).
\]
A phase with $\delta=t\log W$ has $O(\delta^{-2}\log W)$ iterations.
Summing \cref{lem:dynamic-step} over the phases
and using \cref{lem:streaming-support} gives
$O(\mu(G)\eps^{-4}\log W)$ auxiliary updates and
$O(\mu(G)\eps^{-5}\log W+\mu(G)\eps^{-4}\log^2 W)$ local operations,
including phase-end restorations and the final matching computation.
Since the corresponding $U_b$ bound is $\Omega(1/\eps)$,
the amortized computation time is
\[
    O(\mu_w(G))+O(\mu(G)\eps^{-4}\log^2 W)\cdot
    U_b\left(O(nW\eps^{-1}),O((m+n)W\eps^{-1}),\eps/(4C)\right).
\]
The maintenance and computation bounds are amortized jointly
over input updates and MWM computations.

Apply \cref{lem:dynamic-periodic-rebuilding}
using the computation above at accuracy $\eps/2$.
After saving a matching of weight $s$,
maintain its surviving edges and
rebuild on the $K=\lceil\eps s/(8W)\rceil$-th subsequent update
before answering it, taking $K=1$ if $s=0$.
At the next rebuild,
$\mu_w(G)\leq s/(1-\eps/2)+KW=O(KW/\eps)$.
Charging the computation cost over these $K$ updates
gives the stated bound,
including maintenance and explicit matching replacement.
\end{proof}

\subsection{Removing the weight dependence}
\label{sec:dynamic-weight-reduction}
We apply the following weight reduction
to the dynamic algorithm of \cref{lem:dynamic-periodic-bound}.
It reduces the required weight range to $O(1/\eps)$
with $\log^{O(1)}(1/\eps)$ multiplicative overhead.

\begin{lemma}[{\cite[Theorem~4.8]{Chen26}}]
\label{lem:dynamic-weight-reduction}
Suppose that, for every $R\geq1$ and $0<\rho\leq1/2$,
there is a fully dynamic algorithm that,
on an initially empty general graph with at most $n$ vertices and $m$ edges
and weights in $[1,R]$,
explicitly maintains a $(1-\rho)$-MWM
in amortized update time $U(n,m,R,\rho)$.

Then, for every $0<\eps\leq1/2$,
there is a fully dynamic algorithm that,
on an initially empty general graph with at most $n$ vertices and $m$ edges
and weights in $[1,W]$,
explicitly maintains a $(1-\eps)$-MWM
in amortized update time
\[
    \otilde\!\left(
        U\!\left(n,m,O(\eps^{-1}),
            \Theta\!\left(\frac{\eps}{\log^2(1/\eps)}\right)\right)
        +\eps^{-3}\right).
\]
The reduction preserves determinism.
\end{lemma}

\begin{samepage}
\begin{theorem}
\label{thm:dynamic-basic}
Given a bipartite unweighted matching algorithm
with update bound $U_b(n,m,\rho)$ as above, for every $0<\eps\leq1/4$,
there is a fully dynamic algorithm that,
on an initially empty general graph with at most $n$ vertices and $m$ edges
and weights in $[1,W]$,
explicitly maintains a $(1-\eps)$-MWM in amortized update time
\[
    \otilde\!\left(\eps^{-6}\cdot
        U_b\!\left(O(n\eps^{-2}\log^2(1/\eps)),
            O((m+n)\eps^{-2}\log^2(1/\eps)),
            \Theta\!\left(\frac{\eps}{\log^2(1/\eps)}\right)\right)\right).
\]
The algorithm is deterministic when the bipartite algorithm is deterministic.
\end{theorem}
\end{samepage}

\begin{proof}
Apply \cref{lem:dynamic-weight-reduction}
to \cref{lem:dynamic-periodic-bound}, using weight range $R=O(1/\eps)$
and accuracy $\rho=\Theta(\eps/\log^2(1/\eps))$.
The update-time factor becomes
$R\rho^{-5}\log^2 R=\otilde(\eps^{-6})$,
and the auxiliary graph sizes are multiplied by
$R/\rho=O(\eps^{-2}\log^2(1/\eps))$.
The bipartite accuracy is $\Theta(\rho)$,
giving the stated arguments of $U_b$.
The weight reduction adds a $\log^{O(1)}(1/\eps)$ multiplicative factor
and $\otilde(\eps^{-3})$ overhead
which is dominated.
Both reductions preserve determinism.
\end{proof}

\subsection{Further acceleration}
\label{sec:dynamic-acceleration}
We improve the reduction overhead of \cref{thm:dynamic-basic}
by two factors of $1/\eps$.
First, we reuse the cut partition across optimization iterations,
using its remaining odd-set slack to support subsequent crossing updates.
This reduces the cost of reconstructing the partition
and changing internal-edge copies in the auxiliary graphs,
saving one factor of $1/\eps$.

Second, we adapt the matching-composition argument of \cite{Chen26}
to combine matchings computed at different times.
This permits less frequent recomputation and saves another factor of $1/\eps$.
The composition lemma applies independently of our optimization framework.

\subsubsection{Reusing cut partitions}
\label{subsec:dynamic-cut-reuse}
The basic algorithm reconstructs the cut partition at every oracle call.
We instead reuse the partition and its internal graph across calls within a phase.

Modify the partition construction and internal-edge updates
in \cref{alg:dynamic-step} as follows.
At the first oracle call of a phase, construct $\Q,H$ and set $b=0$.
After accepting a crossing update, add $\eta$ to $b$.
Before an oracle call with $b+\eta>\delta/8$,
reconstruct $\Q,H$ and reset $b=0$.
Otherwise, retain $\Q,H$ and omit their construction and internal-edge updates.

\begin{lemma}
\label{lem:cut-certificate}
Let $0<\delta\leq1/32$ and $x\in\P_{G,\delta}$,
and let $\Q$ be the partition produced from $x$
by \cref{lem:cut-partition}.
Keep $\Q$ fixed, start with $z=x$,
and allow the following updates:
\begin{enumerate}[label=\arabic*.,nosep]
    \item $z\gets z+\eta\vone_J$, where $\eta\geq0$
    and $J$ is a matching whose edges join distinct parts of $\Q$,
    provided $z_v+\eta\leq1-2\delta$ at every vertex matched by $J$.
    \item $z\gets(1-\eta)z+\eta(1-2\delta)\vone_I$,
    where $0\leq\eta\leq1$ and $I$ is any matching of $G$.
\end{enumerate}
Let $b$ be the sum of the step sizes in updates of the first type.
If $b\leq\delta/8$ throughout the sequence,
every resulting $z$ belongs to $\P_{G,\delta}$.
\end{lemma}

\begin{proof}
For every $S\in\O_G$ with $|S|\leq1/\delta$
and $Q\in\Q$ with $S\cap Q\in\O_G$,
put $U=S\setminus Q$.
We maintain the invariant
$z(S)\leq (|S|-1)/2-(\delta/4-b)|U|$.
The proof of \cref{lem:odd-set-preservation} gives
$x(S)\leq(|S|-1)/2-\delta|U|/4$.
Thus the invariant holds initially.

For $z\gets z+\eta\vone_J$,
the matching $J$ has no edge inside $S\cap Q$.
Each edge of $J\cap E[S]$ therefore uses a distinct vertex of $U$,
so $z(S)$ increases by at most $\eta|U|$.
This preserves the invariant with $b$ replaced by $b+\eta$.

For the second type of update, put $p=(1-2\delta)\vone_I$.
Since $I$ is a matching and $S\cap Q$ is nonempty,
$p(S)\leq(1-2\delta)(|S|-1)/2\leq(|S|-1)/2-\delta|U|$.
Both $z$ and $p$ satisfy the invariant at the current $b$,
so their convex combination does as well.

Nonnegativity and the degree bounds are preserved by both updates.
Every odd set $S$ has a part $Q$ with $S\cap Q\in\O_G$,
so the invariant proves feasibility of all small odd-set constraints.
For $|S|>1/\delta$, the degree bounds give
$z(S)\leq(1-2\delta)|S|/2<(|S|-1)/2$.
\end{proof}

The certificate is also preserved when $z$ is replaced by $z'$ with $0\leq z'\leq z$,
since its invariant and degree constraints are upper bounds.
Thus the arithmetic implementation in \cref{sec:streaming-eval}
also permits partition reuse.

\Cref{lem:cut-certificate} makes the partition reuse described above feasible.
The resulting algorithm improves the runtime of
\cref{lem:dynamic-periodic-bound} by a factor of $1/\eps$,
up to $\log^{O(1)}(1/\eps)$ factors.

\begin{lemma}
\label{lem:dynamic-mwm}
Given a bipartite unweighted matching algorithm
with update bound $U_b(n,m,\rho)$ as above, for every $0<\eps\leq1/4$,
there is a fully dynamic algorithm that,
on an initially empty general graph with at most $n$ vertices and $m$ edges
and weights in $[1,W]$,
explicitly maintains a $(1-\eps)$-MWM in amortized update time
\[
    O\!\left(W\eps^{-4}\log W\log(W/\eps)\cdot
        U_b\!\left(O(nW\eps^{-1}),O((m+n)W\eps^{-1}),\Theta(\eps)\right)\right).
\]
The algorithm is deterministic when the bipartite algorithm is deterministic.
\end{lemma}

\begin{proof}
Use \cref{alg:dynamic-mwm} with the cut-partition reuse described above.
The first-order guarantee of \cref{lem:matching-gain-decomposition}
holds for every $H\subseteq G$.
The inequalities $-\grad F_t(x)_e>0$ for $e\in J$ ensure the degree bounds,
and \cref{lem:cut-certificate} ensures odd-set feasibility.
Thus the approximation and auxiliary maintenance arguments
from \cref{lem:dynamic-periodic-bound} still apply.

We first bound the cost of computing a fresh matching while $G$ is fixed.
In a phase with $\delta=t\log W$, the step size is
$\eta=\Theta(\delta^2/\log W)$.
Reaching $b+\eta>\delta/8$
requires $\Omega(\delta/\eta)$ crossing matching updates.
By \cref{lem:centering}, the phase has $O(1/\eta)$ iterations.
Consequently, the partition is constructed $O(1/\delta)$ times.

At each construction, we compute $\Q,H$,
restore the previous internal edges in $\D^\Delta$,
and remove the new internal edges.
By the proof of \cref{lem:dynamic-step}, these operations require
$O(\mu(G)\delta^{-2})$ auxiliary edge updates
and $O(|\supp(x)|+\mu(G)\delta^{-3})$ local operations.
The constructions therefore use $O(\mu(G)\delta^{-3})$
auxiliary edge updates per phase.
Restoring the final internal graph at the end of the phase uses only
$O(\mu(G)\delta^{-2})$ additional auxiliary edge updates.
Since $\delta$ halves between phases and ends at $\Theta(\eps)$,
these costs sum to $O(\mu(G)\eps^{-3})$ auxiliary edge updates.

The remaining oracle operations are performed at every call,
even when $\Q,H$ are retained.
Setting and resetting the rounded duals uses
$O(\mu(G)\delta^{-1})$ auxiliary edge updates,
and querying the crossing matching takes $O(\mu(G)\delta^{-1})$ local time,
by the proof of \cref{lem:dynamic-step}.
There are at most $O(\delta^{-2}\log W)$ calls in a phase,
so the dual updates contribute $O(\mu(G)\delta^{-3}\log W)$
auxiliary edge updates per phase.
Summing over phases, including the internal-edge updates above,
gives $O(\mu(G)\eps^{-3}\log W)$ auxiliary edge updates in total.

For the local work, every iteration makes $O(1)$ support scans.
The support bound of \cref{lem:streaming-support}
and the $O(\eps^{-2}\log W)$ total iteration bound
therefore give $O(\mu(G)\eps^{-4}\log^2 W)$ time for these scans.

Since the weights $\widehat r$ change,
each oracle call recomputes $I$ in
$O(\mu(G)\delta^{-2}\log(1/\delta))$ time.
Over all phases, these computations take
$O(\mu(G)\eps^{-4}\log W\log(1/\eps))$ time.
The other partition-construction work sums to $O(\mu(G)\eps^{-4})$.
Dual computation from \cref{lem:streaming-eval},
crossing queries, and the final matching computation have smaller costs;
thus the total local work is
$O(\mu(G)\eps^{-4}\log W\log(W/\eps))$.

Every maintained instance has $O(nW\eps^{-1})$ vertices
and $O((m+n)W\eps^{-1})$ edges,
since $\Delta_\ell\geq\Delta_0=\Theta(\eps)$,
and uses bipartite accuracy $\eps/(4C)$.
Multiplying the auxiliary-update count by the corresponding $U_b$ bound
accounts for all temporary updates.
The lower bound $U_b=\Omega(1/\eps)$
also absorbs the local work into the bound below.
The two zero-dual queries for $L,r$ add $O(\mu_w(G))$ time,
with $\mu(G)\leq r\leq3\mu(G)$ as before.
Hence computing a fresh $(1-\eps)$-MWM takes amortized time
\[
    O\!\left(\mu_w(G)+\mu(G)\eps^{-3}\log W\log(W/\eps)\cdot
        U_b\!\left(O(nW\eps^{-1}),O((m+n)W\eps^{-1}),\eps/(4C)\right)\right).
\]

The auxiliary maintenance cost is unchanged
from the proof of \cref{lem:dynamic-periodic-bound}.
Apply \cref{lem:dynamic-periodic-rebuilding}
to computations at accuracy $\eps/2$.
The charging argument in the proof of \cref{lem:dynamic-periodic-bound}
multiplies the recomputation factor by $O(W/\eps)$,
giving the stated update bound, including auxiliary maintenance
and explicit matching replacement.
\end{proof}

\begin{samepage}
\begin{theorem}
\label{thm:dynamic-baseline}
Given a bipartite unweighted matching algorithm
with update bound $U_b(n,m,\rho)$ as above, for every $0<\eps\leq1/4$,
there is a fully dynamic algorithm that,
on an initially empty general graph with at most $n$ vertices and $m$ edges
and weights in $[1,W]$,
explicitly maintains a $(1-\eps)$-MWM in amortized update time
\[
    \otilde\!\left(\eps^{-5}\cdot
        U_b\!\left(O(n\eps^{-2}\log^2(1/\eps)),
            O((m+n)\eps^{-2}\log^2(1/\eps)),
            \Theta\!\left(\frac{\eps}{\log^2(1/\eps)}\right)\right)\right).
\]
The algorithm is deterministic when the bipartite algorithm is deterministic.
\end{theorem}
\end{samepage}

\begin{proof}
Apply \cref{lem:dynamic-weight-reduction}
to \cref{lem:dynamic-mwm} with $R=O(1/\eps)$
and $\rho=\Theta(\eps/\log^2(1/\eps))$.
The multiplicative factor becomes
$R\rho^{-4}\log R\log(R/\rho)=\otilde(\eps^{-5})$.
The remaining substitutions and absorption of the additive overhead
are the same as in the proof of \cref{thm:dynamic-basic}.
\end{proof}

\subsubsection{Customizing the weight reduction framework}
\label{subsec:dynamic-temporal-composition}
To improve the black-box application of \cref{lem:dynamic-weight-reduction},
we revisit the matching-composition argument of \cite{Chen26}.
It uses overlapping weight windows,
each consisting of all edges whose weights lie in an interval.
With chosen intervals of weight ratio $O(1/\eps)$,
the union of sufficiently accurate approximate MWMs of these windows
contains a $(1-\eps)$-MWM of the whole graph
\cite[Lemma~2.4]{Chen26}.

For a weighted graph $G=(V,E,w)$,
let $E_i=\{e\in E:2^i\leq w_e<2^{i+1}\}$ be \emph{weight class} $i$.
Write $G_{[i,j]}$ for the subgraph with edge set $\bigcup_{k=i}^j E_k$,
and $G_{\geq i}$ for the subgraph with edge set $\bigcup_{k\geq i}E_k$.

The composition guarantee requires each local matching
to approximate its current window.
To allow less frequent recomputation,
we bound the loss in their union directly in terms of
the edge updates.
In the dynamic application, $G$ denotes the current graph.
For an integer $h\geq1$, we will consider the windows $G_{[i,i+h]}$.

We will use periodic recomputation for each window $i$.
Let $G^{(i)}$ denote the entire input graph
at the last recomputation for window $i$.
The subgraph $G^{(i)}_{[i,i+h]}$ is that window at the time of the computation.
Let $M_i$ be a $(1,\beta_i)$-MWM of it.
Subsequent updates may change the window optimum and delete edges of $M_i$.
Let $D_i$ consist of the weighted edges inserted or deleted in this window
since that computation, identifying each edge by its endpoints and weight.

Deleting $D_i$ from $G^{(i)}_{[i,i+h]}$ and $G_{[i,i+h]}$ makes them identical.
Consequently, $M_i\setminus D_i\subseteq M_i\cap G$.
Let $H=\bigcup_i(M_i\cap G)$ be the union of their edges present in $G$.
The following lemma bounds how much matching weight is preserved in $H$.

\begin{lemma}
\label{lem:temporal-window-composition}
With the notation above,
set $\ell=\lfloor\log W\rfloor$, and fix an integer $h\geq1$.
Suppose that, for each $0\leq i\leq\ell$,
$M_i$ is a $(1,\beta_i)$-MWM of $G^{(i)}_{[i,i+h]}$ and
$G^{(i)}_{[i,i+h]}\setminus D_i=G_{[i,i+h]}\setminus D_i$.
Then $H$ satisfies
\[
    \mu_w(H)
    \geq\bigl(1-O(2^{-h})\bigr)\mu_w(G)
        -\sum_{i=0}^{\ell}\beta_i
        -O\!\left(\sum_{i=0}^{\ell}2^i|D_i|\right).
\]
\end{lemma}

\begin{proof}
Write $H_{\geq i}=H\cap G_{\geq i}$
for the subgraph of $H$ containing edges in classes $i$ and higher.
We compare the increase in optimum weight when class $i$ is included
in $H$ with the corresponding increase in $G$.
Summing these comparisons will give the desired bound.

For each $0\leq i\leq\ell$, set
$\lambda_i=2^{i+2}(|D_i|+\mu(G_{\geq i+h+1}))$.
This term will bound the weight discarded in the exchange construction below.

\begin{claim}
\label{clm:temporal-window-exchange}
For each $0\leq i\leq\ell$,
\[
    \mu_w(H_{\geq i})-\mu_w(H_{\geq i+1})
    \geq\mu_w(G_{\geq i})-\mu_w(G_{\geq i+1})-\beta_i-2\lambda_i.
\]
\end{claim}

\begin{proof}[Proof of \cref{clm:temporal-window-exchange}]
Fix $i$ and let $M'$ be an MWM of $H_{\geq i+1}$.
We adapt the alternating component construction
from the proof of \cite[Lemma~2.1]{Chen26}
to the suffix windows and changed edges $D_i$.
We use the following construction twice.
In each application, we obtain two new matchings
whose total weight is at least that of the two starting matchings minus $\lambda_i$:
\begin{enumerate}[label=(\roman*),nosep]
    \item Starting from $M_i$ and $M'$, construct a new matching
    in $H_{\geq i}$ and another in $G^{(i)}_{[i+1,i+h]}$.
    \item Starting from an MWM of $G_{\geq i}$
    and an MWM of $G^{(i)}_{[i+1,i+h]}$, construct a new matching
    in $G^{(i)}_{[i,i+h]}$ and another in $G_{\geq i+1}$.
\end{enumerate}

Common edges can be kept in both resulting matchings.
The remaining edges form alternating paths and even cycles.
Delete every class-$i$ edge in $D_i$.
From each remaining edge in $D_i$ or above class $i+h$,
select the first class-$i$ edge in each direction along its component, if any,
and then delete all selected edges.
All deleted edges have class $i$;
the edges in classes $i+h+1$ and higher are kept.
After these deletions, no path connects a surviving class-$i$ edge
to an edge in $D_i$ or above class $i+h$.
Thus a component containing a class-$i$ edge
contains only edges in classes $i,\ldots,i+h$ outside $D_i$.
The window graphs are identical after deleting $D_i$,
so every edge of this component is present in both
$G^{(i)}_{[i,i+h]}$ and $G_{[i,i+h]}$.
On these components, send the first starting matching's edges
to the first resulting matching
and the second starting matching's edges to the second resulting matching;
on all other components, reverse the assignment.
The latter contain no class-$i$ edges,
so the reversed assignment also meets the requirements in (i) and (ii).

To bound the loss, charge each selected class-$i$ edge
to the edge from which it was selected,
and charge each deleted class-$i$ edge in $D_i$ to itself.
Each edge in $D_i$ or in classes $i+h+1$ and higher
receives at most two charges, each of weight less than $2^{i+1}$.
The edges in classes $i+h+1$ and higher belong to $M'$ in (i)
and to the MWM of $G_{\geq i}$ in (ii).
Their number is therefore at most $\mu(G_{\geq i+h+1})$,
so each exchange loses at most $\lambda_i$.

The two applications give, respectively,
\begin{enumerate}[label=(\roman*),nosep]
    \item $\mu_w(H_{\geq i})+\mu_w(G^{(i)}_{[i+1,i+h]})
    \geq w(M_i)+w(M')-\lambda_i$
    \item $\mu_w(G^{(i)}_{[i,i+h]})+\mu_w(G_{\geq i+1})
    \geq\mu_w(G_{\geq i})+\mu_w(G^{(i)}_{[i+1,i+h]})-\lambda_i$
\end{enumerate}
Add the inequalities and use $w(M')=\mu_w(H_{\geq i+1})$ and
$w(M_i)\geq\mu_w(G^{(i)}_{[i,i+h]})-\beta_i$.
The terms $\mu_w(G^{(i)}_{[i,i+h]})$ and $\mu_w(G^{(i)}_{[i+1,i+h]})$
cancel, leaving the claimed inequality.
\end{proof}

Sum \cref{clm:temporal-window-exchange} over $i$.
The differences telescope to $\mu_w(H)$ and $\mu_w(G)$,
respectively, giving
\[
    \mu_w(G)-\mu_w(H)
    \leq\sum_i\beta_i+8\sum_i2^i|D_i|
        +8\sum_i2^i\mu(G_{\geq i+h+1}).
\]

The first two sums give the stated errors from local approximation and changed edges.
We finish by bounding the last sum, which comes from edges above the windows.
For each $j$, partition an MCM of $G_{\geq j}$ by weight class.
For every $k\geq j$, its edges in class $k$ form a matching in $G_{[k,k]}$,
and each has weight at least $2^k$.
This class therefore contributes at most $2^{-k}\mu_w(G_{[k,k]})$ edges, giving
\[
    2^j\mu(G_{\geq j})
    \leq\sum_{k\geq j}2^{j-k}\mu_w(G_{[k,k]}).
\]
Summing over $j$ uses $\sum_{j=0}^k2^{j-k}<2$ for the first inequality below;
the second is \cite[Lemma~2.2]{Chen26} with windows containing one weight class:
\begin{equation}
    \sum_{j=0}^{\ell}2^j\mu(G_{\geq j})
    \leq2\sum_{j=0}^{\ell}\mu_w(G_{[j,j]})
    \leq10\mu_w(G).
    \label{eq:dynamic-suffix-sum}
\end{equation}
Substituting $j=i+h+1$ therefore gives
\[
    8\sum_{i=0}^{\ell}2^i\mu(G_{\geq i+h+1})
    \leq40\cdot2^{-h}\mu_w(G).
\]
Combining these error bounds proves the lemma.
\end{proof}

\Cref{lem:temporal-window-composition} allows each window matching
to be recomputed on its own schedule.
Set $h=\lceil\log(1/(c_0\eps))\rceil$ and $\rho=c_0\eps/(h+1)$
for a sufficiently small constant $c_0>0$.
Given a cardinality estimate $\widehat r_i$ of window $i$,
compute a $(1-\rho)$-MWM $M_i$ and recompute it after
$K_i=\max\{1,\lceil c_0\eps\widehat r_i\rceil\}$ updates to that window.
Maintain an explicit $(1-\eps/2)$-MWM of $H$.

\begin{lemma}
\label{lem:dynamic-temporal-reduction}
Given a bipartite unweighted matching algorithm
with update bound $U_b(n,m,\rho)$ as above, for every $0<\eps\leq1/4$,
there is a fully dynamic algorithm that,
on an initially empty general graph with at most $n$ vertices and $m$ edges
and weights in $[1,W]$,
explicitly maintains a $(1-\eps)$-MWM in amortized update time
\[
    \otilde\left(\eps^{-4}\right)\cdot\left(
        U_b\left(O(nW\eps^{-1}\log(1/\eps)),
    O((m+n)W\eps^{-1}\log(1/\eps)),
        \Theta\left(\frac{\eps}{\log(1/\eps)}\right)\right)
        +\log W\right).
\]
The algorithm is deterministic when the bipartite algorithm is deterministic.
\end{lemma}

\begin{proof}
Use the MWM computation from the proof of \cref{lem:dynamic-mwm}
on each window after multiplying its weights by $2^{-i}$.
Use the weight bound $\max\{8,\min\{2^{h+1},W/2^i\}\}$,
which is both $O(1/\eps)$ and at most $W$.
The same computation gives an estimate
$\mu(G_{[i,i+h]})\leq\widehat r_i\leq3\mu(G_{[i,i+h]})$.

We first bound the approximation loss between rebuilds.
For each window, let $q_i$ count its updates since the last rebuild,
$G^{(i)}$ be the input graph at that rebuild,
and $D_i$ the weighted edges inserted or deleted in the window since then.
The old and current windows become identical after removing $D_i$,
and $|D_i|\leq q_i$.
The local approximation error is
$\beta_i=\rho\mu_w(G^{(i)}_{[i,i+h]})$, measured in the original weights.
By \cref{lem:temporal-window-composition},
\[
    \mu_w(G)-\mu_w(H)
    \leq O(2^{-h})\mu_w(G)+\sum_i\beta_i
        +O\!\left(\sum_i2^iq_i\right).
\]

The period length bounds the changed-edge loss.
Write $r_i^{\rm old}$ for the window's maximum matching cardinality
at its last rebuild.
Since $q_i<K_i$, we have $q_i\leq3c_0\eps r_i^{\rm old}$.
Also, $\mu(G_{[i,i+h]})\geq r_i^{\rm old}-q_i$,
so $q_i=O(c_0\eps\mu(G_{[i,i+h]}))$.
Since $G_{[i,i+h]}\subseteq G_{\geq i}$,
\eqref{eq:dynamic-suffix-sum} gives
$\sum_i2^iq_i=O(c_0\eps\mu_w(G))$.

For the local approximation loss,
each update changes the window optimum by less than $2^{i+h+1}$.
Together with
$\sum_i\mu_w(G_{[i,i+h]})=O((h+1)\mu_w(G))$
from \cite[Lemma~2.2]{Chen26}, this gives
\[
    \sum_i\beta_i
    \leq\rho\sum_i\bigl(\mu_w(G_{[i,i+h]})+2^{i+h+1}q_i\bigr)
    =O\bigl(\rho(h+1)+\rho2^hc_0\eps\bigr)\mu_w(G).
\]
Since $2^{-h}\leq c_0\eps$, $\rho(h+1)=c_0\eps$,
and $\rho2^hc_0\eps=O(c_0\eps)$,
the total loss is $O(c_0\eps\mu_w(G))$.
Choosing $c_0$ sufficiently small gives
$\mu_w(H)\geq(1-\eps/4)\mu_w(G)$.
Thus the maintained $(1-\eps/2)$-MWM of $H$ is a $(1-\eps)$-MWM of $G$.

We next bound the update time.
Each auxiliary instance has $O(nW/\rho)$ vertices,
$O((m+n)W/\rho)$ edges, and bipartite accuracy $\Theta(\rho)$.
Since the normalized window has weight bound $O(1/\eps)$,
the proof of \cref{lem:dynamic-mwm} bounds its recomputation cost by
$\otilde(\mu(G_{[i,i+h]})\rho^{-3})$
times the corresponding $U_b$ bound.
The normalized $O(\mu_w)$ work in that proof is absorbed
using $U_b=\Omega(1/\rho)$.

Charge each rebuild to the preceding $K_i$ updates.
The window cardinality stays $O(K_i/\eps)$ throughout this period,
by the choice of $K_i$, since each update changes cardinality by at most one.
Rebuilding therefore costs $\otilde(\rho^{-3}\eps^{-1})$
times the same $U_b$ bound per window update.
Auxiliary maintenance adds only $O(\eps^{-1}\rho^{-1})$ times this bound.
Each input update affects at most $h+1=O(\log(1/\eps))$ windows,
so $\rho=\Theta(\eps/\log(1/\eps))$ gives
$\otilde(\eps^{-4})$ times the stated $U_b$ bound.
Joint amortization follows from the proof of \cref{lem:dynamic-mwm}.

Replacing a window matching changes $O(K_i/\eps)$ edges in $H$.
Between rebuilds, each window update changes at most one edge of $M_i\cap G$.
Thus $H$ receives $O((h+1)/\eps)$ amortized edge updates per input update,
with constant bookkeeping time per changed edge occurrence.
Its maximum vertex degree is at most $\ell+1$,
so maintaining its matching costs
$\otilde((\ell+1)\eps^{-3})$ amortized time per change,
by \cite[Corollary~4.14]{Chen26}.
This contributes $\otilde(\eps^{-4}\log W)$ time per input update,
giving the remaining term in the statement.
\end{proof}

\begin{proof}[Proof of \cref{thm:dynamic-reduction}]
Apply \cref{lem:dynamic-weight-reduction}
to \cref{lem:dynamic-temporal-reduction},
using weight bound $R=O(1/\eps)$
and accuracy $\beta=\Theta(\eps/\log^2(1/\eps))$.
The update-time overhead is
$\beta^{-4}\log^{O(1)}(1/\beta)=\otilde(\eps^{-4})$.
The auxiliary size multiplier is
$R\beta^{-1}\log(1/\beta)=O(\eps^{-2}\log^3(1/\eps))$,
and the bipartite accuracy is\\
$\Theta(\beta/\log(1/\beta))=\Theta(\eps/\log^3(1/\eps))$.
These give the stated arguments of $U_b$.
The union-matching term has $\log R=O(\log(1/\eps))$
and is absorbed into the $\log^{O(1)}(1/\eps)$ factor since $U_b\geq1$.
The outer weight reduction adds a $\log^{O(1)}(1/\eps)$ factor
and $\otilde(\eps^{-3})$ time per update,
which are absorbed as in \cref{thm:dynamic-basic}.
Both reductions preserve determinism.
\end{proof}

\section{Concave Objectives and Decremental General Matching}
\label{sec:decremental}
In the decremental matching problem,
we start with an initial graph and
maintain an explicit $(1-\eps)$-MWM
under edge deletions.

\begin{samepage}
\decrementaltheorem*
\end{samepage}

We follow the framework of \cite{ChenST25},
which repeatedly rebuilds a primal entropy-regularized fractional matching
after deletions remove a $\Theta(\eps)$ fraction of its weight.
We use their rebuilding guarantee as a black box.
To compute these rebuilds, \cref{sec:decremental-rebuild}
extends the optimization framework of \cref{sec:optimization}
to concave objectives.

The implementation in \cref{sec:decremental-oracles}
reuses the cut partition of \cref{sec:cut-partition}
and the split into internal and crossing directions.
Accurate optimization is confined to the internal graph,
which has $O(n/\delta)$ edges.
On crossing edges, a constant approximation suffices.
Both problems maximize separable concave objectives over matching polytopes.
We use \cref{lem:cst-concave-matching} to compute the directions.

\subsection{Primal entropy regularization under adversarial deletions}
\label{sec:decremental-robustness}
An adversary may delete a large fraction of the current matching
while leaving the maximum matching weight almost unchanged.
The key idea of \cite{ChenST25}
is to use primal entropy regularization to
spread out the fractional matching mass.
Deleting enough weight from an approximate maximizer
forces the optimum regularized value to decrease.
This decrease bounds the number of rebuilds,
even when deletions are chosen after observing the maintained matching.

We use the rebuilding procedure of
\cite[Algorithms~1--2 and Theorem~3.5]{ChenST25}.
Fix $\lambda=\Theta(\eps/\log n)$ with a sufficiently small constant.
Each rebuild uses their primal entropy objective
$\Psi(x)=\sum_{e\in E}\psi_e(x_e)$,
where $\psi_e(s)=w_es+\lambda w_es\ln(\gamma/(w_es))$ and $\psi_e(0)=0$.
The procedure chooses the scale $\mu_w(G)\leq\gamma\leq nW$.
Write $\Psi_G^*=\max_{x\in\P_G}\Psi(x)$.
The entropy bound in \cite[Lemma~4.1]{ChenST25} gives
$0\leq\Psi(x)-w^\top x\leq O(\lambda\log n)\mu_w(G)$ on $\P_G$,
so $\mu_w(G)\leq\Psi_G^*\leq2\mu_w(G)$.

\begin{lemma}[{\cite[Theorem~3.5]{ChenST25}}]
\label{lem:cst-decremental}
The procedure above maintains a fractional $(1-\eps)$-MWM
using $O(\eps^{-2}\log^2 n)$ calls that compute $x\in\P_G$ with
$\Psi_G^*-\Psi(x)=O(\lambda\eps^2\Psi_G^*)$,
where the implied constant is sufficiently small.
\end{lemma}

\subsection{Concave objective optimization}
\label{sec:decremental-rebuild}
To compute the approximate maximizers required by
\cref{lem:cst-decremental},
we extend \cref{sec:optimization} from maximizing $w^\top x$
to maximizing any continuous concave objective $\Psi$ on $\P_G$
with $\Psi(0)=0$ and $\Psi_G^*\geq\mu(G)$,
where $\Psi_G^*=\max_{x\in\P_G}\Psi(x)$.
We minimize the corresponding regularized objective
$\Phi_t(x)=-\Psi(x)+t(\log e)\sum_v y_{t,v}(x)$,
and let $\Phi_t^*=\min_{z\in\P_G}\Phi_t(z)$.
For $0<t\leq1/(32\log W)$,
set $\delta=t\log W$ as in \cref{sec:local-optimization},
and let $\eta=\Theta(t\delta)$ be a power of two
with a sufficiently small implied constant.
The following analogue of \cref{lem:potential-bias}
converts a small gap in $\Phi_t$ into a small gap in $\Psi$.

\begin{lemma}
\label{lem:entropy-potential-bias}
For every $x\in\P_G$,
$\Psi_G^*-\Psi(x)\leq\Phi_t(x)-\Phi_t^*
+O(\delta\Psi_G^*)$.
\end{lemma}

\begin{proof}
Let $z^*$ maximize $\Psi$ on $\P_G$ and put $z=(1-4\delta)z^*$.
The degree $z_v$ is a convex combination
of $0$ and $1-4\delta$,
with coefficients $1-z_v^*$ and $z_v^*$.
Since the vertex dual is convex in the degree
and equals $1$ at degree $1-4\delta$,
\[
 y_{t,v}(z)
 \leq(1-z_v^*)y_{t,v}(0)+z_v^*
 \leq y_{t,v}(x)+z_v^*.
\]
Since $z^*$ is a fractional matching,
$\sum_vz_v^*=2\sum_ez_e^*\leq2\mu(G)\leq2\Psi_G^*$.
The definition of $\Phi_t$ and $\Phi_t(z)\geq\Phi_t^*$ therefore give
$\Psi(z)-\Psi(x)\leq\Phi_t(x)-\Phi_t^*+2t(\log e)\Psi_G^*$.
Concavity and $\Psi(0)=0$ give
$\Psi(z)\geq(1-4\delta)\Psi_G^*$.
Combining these bounds with $t\log e=O(\delta)$ proves the claim.
\end{proof}

For a direction $d$,
the first-order change of the vertex-dual term is
$t(\log e)\sum_v\grad y_{t,v}(x)^\top d=y_t(x)^\top Ad$.
We combine this with the finite difference of $-\Psi$ and get
\[
 \mathcal L_x(d)\defeq y_t(x)^\top Ad-
                  \frac{\Psi(x+\eta d)-\Psi(x)}\eta.
\]
It is defined when $x+\eta d\in\P_G$,
equals $\grad F_t(x)^\top d$ for the linear objective $\Psi(z)=w^\top z$,
and does not require $\Psi$ to be differentiable.
The following analogue of \cref{lem:shrinking-curvature}
bounds the change in $\Phi_t$.

\begin{lemma}
\label{lem:entropy-curvature}
Let $x\in\P_{G,\delta}$.
\begin{enumerate}[label=\arabic*.,nosep]
    \item If $\mathcal L_x(-x)\leq-\delta \Psi_G^*/8$, then
    $\Phi_t(x-\eta x)\leq\Phi_t(x)+\frac{\eta}{2}\mathcal L_x(-x)$.
    \item If $\mathcal L_x(-x)\geq-2\delta \Psi_G^*$, then
    $\Phi_t(x+\eta d)\leq\Phi_t(x)+\eta\mathcal L_x(d)
    +O(\eta^2\Psi_G^*/t)$
    for every $d$ with $x+\eta d\in\P_G$,
    $\|Ad\|_\infty\leq1$, and $\|Ad\|_1\leq4\mu(G)$.
\end{enumerate}
\end{lemma}

\begin{proof}
Concavity and $\Psi(0)=0$ give
$[\Psi(x)-\Psi((1-\eta)x)]/\eta\leq\Psi(x)\leq\Psi_G^*$,
so the definition of $\mathcal L_x$ gives
$\sum_vx_vy_{t,v}(x)\leq\Psi_G^*-\mathcal L_x(-x)$.

For the first claim,
the Taylor calculation in the proof of \cref{lem:shrinking-curvature} gives
$\Phi_t(x-\eta x)-\Phi_t(x)
\leq\eta\mathcal L_x(-x)+O(\eta^2/t)\sum_vx_vy_{t,v}(x)$.
Under the hypothesis, the sum is $O(-\mathcal L_x(-x)/\delta)$,
so the small constant in $\eta=\Theta(t\delta)$
bounds the remainder by $-\eta\mathcal L_x(-x)/2$.

For the second claim, the sum is $O(\Psi_G^*)$.
With this bound and $\mu(G)\leq\Psi_G^*$,
the proof of \cref{lem:shrinking-curvature}
bounds the second-order change of the vertex-dual term by $O(\eta^2\Psi_G^*/t)$.
The first-order change and the change in $-\Psi$
sum to $\eta\mathcal L_x(d)$.
\end{proof}

The corresponding oracle guarantees are the following analogues of
\cref{def:first-order-oracle,def:step-oracle}.

\begin{definition}
\label{def:entropy-eval}
Given $x\in\P_{G,\delta}$, $t$, and a direction $d$
with $x+\eta d\in\P_G$,
$\Eval(x,t,d)$ returns $v$ satisfying
$|v-\mathcal L_x(d)|\leq\delta \Psi_G^*/1024$.
\end{definition}

\begin{definition}
\label{def:entropy-step}
Given $x\in\P_{G,\delta}$,
$\Step(x,t)$ returns $d$ satisfying:
\begin{enumerate}[label=\arabic*.,nosep]
    \item $x+\eta d\in\P_{G,\delta}$,
    $\|Ad\|_\infty\leq1$, and $\|Ad\|_1\leq4\mu(G)$.
    \item For an absolute constant $c\geq1$ fixed for the oracle,
    \[
        \mathcal L_x(d)
        \leq\frac1c\min_{d^\prime:\,x+d^\prime\in\P_G}
                     \mathcal L_x(d^\prime)+c\delta \Psi_G^*.
    \]
\end{enumerate}
\end{definition}

Since $\mathcal L_x$ need not be linear in $d$,
we evaluate the shrinking direction $-x$ directly.

\begin{lemma}
\label{lem:decremental-rebuild}
Let $0<\eps\leq1$.
Run \cref{alg:phased-optimization}
with an estimate $\Psi_G^*/4\leq L\leq\Psi_G^*$,
the oracles in \cref{def:entropy-eval,def:entropy-step},
and the shrinking test $\Eval(x,t,-x)\leq-3\delta L/4$
in \cref{alg:fixed-phase}.
It returns $x\in\P_G$ with $\Psi_G^*-\Psi(x)=O(\eps\Psi_G^*)$
in $O(\log W/\eps^2)$ iterations.
\end{lemma}

\begin{proof}
We follow the proofs of
\cref{lem:centering,lem:phase-transition,lem:phased-optimization},
with $\Phi_t$, $\mathcal L_x(d)$, and $\Psi_G^*$
in place of $F_t$, $\grad F_t(x)^\top d$, and $\mu_w(G)$.

Within a phase, the evaluation-error calculation in the proof of \cref{lem:centering}
gives $\mathcal L_x(-x)\leq-\delta\Psi_G^*/8$ when shrinking is selected.
Otherwise, it gives $\mathcal L_x(-x)>-\delta\Psi_G^*$
and $\mathcal L_x(d)=-\Omega(\delta\Psi_G^*)$ for an accepted direction.
\Cref{lem:entropy-curvature} therefore gives a decrease of $\Omega(\eta\delta\Psi_G^*)$
per update.
When the phase returns,
$\mathcal L_x(d)\geq-O(\delta\Psi_G^*)$.
For a minimizer $z$ of $\Phi_t$ over $\P_G$,
concavity gives $\Psi(x+\eta(z-x))\geq(1-\eta)\Psi(x)+\eta\Psi(z)$.
Together with convexity of the vertex-dual term, this gives
$\mathcal L_x(z-x)\leq\Phi_t(z)-\Phi_t(x)$,
so \cref{def:entropy-step} gives $\Phi_t(x)-\Phi_t^*=O(\delta\Psi_G^*)$.

The first phase starts at $x=0$ with gap at most $\Psi_G^*$,
since $y_{t,v}(z)\geq y_{t,v}(0)$ for every $z\in\P_G$.
When $t$ halves, the decomposition in the proof of \cref{lem:phase-transition}
applies to $\Phi_t$.
Its first bracket involves only the vertex-dual term and is nonpositive.
For the second, $\Phi_{t/2}(0)-\Phi_{t/2}^*\leq\Psi_G^*$,
and \cref{lem:entropy-potential-bias} at $x=0$
gives $\Phi_t(0)-\Phi_t^*\geq(1-O(\delta))\Psi_G^*$.
Thus every phase starts with gap $O(\delta\Psi_G^*)$
and has $O(1/\eta)=O(\delta^{-2}\log W)$ iterations.
These bounds sum geometrically,
and the final phase has $\delta=\Theta(\eps)$,
so \cref{lem:entropy-potential-bias} gives $\Psi_G^*-\Psi(x)=O(\eps\Psi_G^*)$.
\end{proof}

\subsection{Decremental algorithm}
\label{sec:decremental-oracles}
The primal entropy objective of \cref{sec:decremental-robustness}
satisfies the assumptions of \cref{sec:decremental-rebuild}.
The optimization through $\delta=\Theta(\lambda\eps^2)$
gives the rebuild accuracy required by \cref{lem:cst-decremental}.
It remains to implement $\Eval$ and $\Step$ for this objective.

\subsubsection{First-order evaluations}
\label{sec:entropy-evaluation}

We implement iterate updates as in \cref{sec:streaming-eval}.
By monotonicity and subadditivity,
rounding loses at most $\sum_e\psi_e(h)\leq mWh(1+\lambda\ln(nW/h))$ in $\Psi$.
Choose $h$ so that this is a sufficiently small fraction
of the $\Omega(\eta\delta\Psi_G^*)$ progress per update
in \cref{lem:decremental-rebuild}.

\begin{lemma}
\label{lem:entropy-eval}
For every queried direction,
$\Eval$ satisfies \cref{def:entropy-eval}.
Each evaluation and iterate update takes $O((m+n)\log^{O(1)}n)$ time.
\end{lemma}

\begin{proof}
Evaluate $\mathcal L_x(d)$ using logarithms and exponentials.
It has $O(m+n)$ scalar terms,
and $\eta$ and $\delta$ are inverse-polynomial.
Thus sufficiently accurate inverse-polynomial evaluations
satisfy \cref{def:entropy-eval}
and take $\log^{O(1)}n$ time per term.
Updating the stored iterate and its vertex degrees takes $O(m+n)$ time.
\end{proof}

\subsubsection{Step oracle}
\label{sec:entropy-step}
Apply \cref{lem:cut-partition} to $x$
and denote by $H$ the internal graph of the resulting partition.
This takes $O(m+n\delta^{-3})$ time,
and $H$ has $O(n/\delta)$ edges.
The following analogue of \cref{lem:matching-gain-decomposition}
justifies optimizing the internal and crossing directions separately
and choosing the better direction.

\begin{lemma}
\label{lem:entropy-direction-decomposition}
Let $x\in\P_{G,\delta}$ and $z\in\P_G$.
Write $z=z_H+z_C$,
where $z_H$ and $z_C$ are the restrictions of $z$ to $H$ and $G\setminus H$.
Then
\[
 \mathcal L_x((1-2\delta)z_H-x)
 +\mathcal L_x(z_C)
 \leq\mathcal L_x(z-x)+O(\delta\Psi_G^*).
\]
\end{lemma}

\begin{proof}
The vertex-dual terms of the left-hand side
differ from those of $\mathcal L_x(z-x)$ by $-2\delta y_t(x)^\top Az_H\leq0$.
It remains to bound
\[
 \Psi((1-\eta)x+\eta z)+\Psi(x)
 -\Psi((1-\eta)x+\eta(1-2\delta)z_H)-\Psi(x+\eta z_C)
\]
by $O(\eta\delta\Psi_G^*)$, edge by edge.
Since $\psi'_e(s)=(1-\lambda)w_e+\lambda w_e\ln(\gamma/(w_es))$
and $\ln(1/(1-\eta))\leq2\eta$,
the contribution of a crossing edge with $x_e>0$ is
\[
 \begin{aligned}
 &\psi_e((1-\eta)x_e+\eta(z_C)_e)-\psi_e((1-\eta)x_e)
       -\psi_e(x_e+\eta(z_C)_e)+\psi_e(x_e)\\
 &\qquad=\lambda w_e\int_0^{\eta(z_C)_e}
          \ln\frac{x_e+s}{(1-\eta)x_e+s}\,ds
 \leq2\lambda w_e\eta^2(z_C)_e.
 \end{aligned}
\]
The contribution is zero when $x_e=0$,
so crossing edges contribute $O(\lambda\eta^2\Psi_G^*)=O(\eta\delta\Psi_G^*)$.
By subadditivity, internal edges contribute at most $\Psi(2\eta\delta z_H)$.
This is $O(\eta\delta\Psi_G^*)$ by the identity
$\Psi(\alpha z)=\alpha\Psi(z)+\lambda\alpha\ln(1/\alpha)w^\top z$
and $\lambda\ln(1/(2\eta\delta))\leq1$.
Dividing by $\eta$ proves the claim.
\end{proof}

\paragraph{The internal direction.}
\label{sec:entropy-internal}

We maximize
$\mathcal L_x(-x)-\mathcal L_x((1-2\delta)z-x)$ over $z\in\P_H$.
This objective is separable and concave, vanishes at $z=0$,
and is bounded above by $\Psi(\eta z)/\eta=O(\Psi_G^*)$.
Thus relative error $\Theta(\delta)$ gives the required additive error
$O(\delta\Psi_G^*)$.
We use \cref{lem:cst-concave-matching}
for the direction implementations below.
For the objectives of both the internal and crossing directions,
approximating the vertex duals on a sufficiently fine inverse-polynomial grid
gives polynomially bounded duals
and changes each objective uniformly by $O(\delta)$.

\begin{lemma}
\label{lem:entropy-internal}
A randomized algorithm computes $z_I\in\P_H$ such that
$d_I=(1-2\delta)z_I-x$ satisfies
\[
 \mathcal L_x(d_I)
 \leq\min_{z\in\P_{H}}\mathcal L_x((1-2\delta)z-x)
       +O(\delta\Psi_G^*),
\]
with high probability,
in $O(n\delta^{-13}\log^{O(1)}n)$ time.
\end{lemma}

\begin{proof}
Write the objective as $\sum_{e\in E(H)}f_e(z_e)$.
For each edge $e$, let $s_e^\star$ be a maximizer of $f_e$ over $[0,1]$.
This maximizing coordinate can be arbitrarily small,
whereas \cref{lem:cst-concave-matching} requires an inverse-polynomial maximizer.
We use $O(\log n)$ bisection steps on $[0,1]$ to compute $a_e$
with $0\leq a_e\leq s_e^\star$
and sufficiently small inverse-polynomial error $s_e^\star-a_e$.
Every positive bisection endpoint $a_e$ is then inverse-polynomial.
Discard edges with $a_e=0$,
and replace $f_e(s)$ by $f_e(\min\{s,a_e\})$ for each remaining edge.
The modified edge objectives satisfy the hypotheses of
\cref{lem:cst-concave-matching},
and the modifications lose at most $O(\delta)$ in the optimum.

Apply \cref{lem:cst-concave-matching} at relative error $\Theta(\delta)$,
cap each returned coordinate at $a_e$,
and discard coordinates below a sufficiently small inverse-polynomial cutoff.
The resulting $z_I$ is feasible and satisfies the claimed approximation guarantee.
The running time follows from $|E(H)|=O(n/\delta)$
and the evaluation cost in \cref{sec:entropy-evaluation}.
\end{proof}

\paragraph{The crossing direction.}
\label{sec:entropy-crossing}
We maximize $-\mathcal L_x(z)$ over $z\in\P_{G\setminus H}$.
Apply the construction in the proof of \cref{lem:entropy-internal}
at relative error $1/4$.
Retain only coordinates at which the original edge objective is certified positive,
incurring another $O(\delta)$ additive error.

\begin{lemma}
\label{lem:entropy-crossing}
A randomized algorithm computes a nonnegative direction $d_C$
supported on $G\setminus H$
with $x+\eta d_C\in\P_{G,\delta}$, $\|Ad_C\|_\infty\leq1$,
$\|Ad_C\|_1\leq4\mu(G)$, and
\[
 \mathcal L_x(d_C)
 \leq\frac12
 \min_{z\in\P_{G\setminus H}}
       \mathcal L_x(z)+O(\delta\Psi_G^*),
\]
with high probability, in $O((m+n)\log^{O(1)}n)$ time.
\end{lemma}

\begin{proof}
The approximation guarantee and running time follow from
\cref{lem:cst-concave-matching} and the $O(\delta)$ loss above.
For every retained crossing edge $e=uv$,
positivity of its edge objective and the inverse-polynomial coordinate bound
give $y_{t,u}(x)+y_{t,v}(x)<2w_e$
for a sufficiently small constant in $\lambda$.
A vertex with $x_v+\eta>1-2\delta$ has
$y_{t,v}(x)>W^2 2^{-\eta/t}>2W$,
so no retained edge is incident to it.
Thus $x+\eta d_C$ satisfies the degree bounds,
and \cref{lem:fractional-crossing} gives the odd-set and norm bounds.
\end{proof}

To implement $\Step$, evaluate $d_I$ and $d_C$ using $\Eval$
and return the better direction $d$.

\begin{lemma}
\label{lem:decremental-step}
$\Step$ satisfies \cref{def:entropy-step}
with high probability in
$O((m+n\delta^{-13})\log^{O(1)}n)$ time.
\end{lemma}

\begin{proof}
Since $x+d_I=(1-2\delta)z_I\in\P_{G,\delta}$,
the norm bounds for $d_I$ follow as in \cref{sec:local-optimization},
and $x+\eta d_I\in\P_{G,\delta}$ follows by convexity.
\Cref{lem:entropy-crossing} gives these guarantees for $d_C$.

For $z\in\P_G$, applying \cref{lem:entropy-internal,lem:entropy-crossing}
to its restrictions $z_H$ and $z_C$,
and then \cref{lem:entropy-direction-decomposition}, gives
$\mathcal L_x(d_I)+2\mathcal L_x(d_C)\leq\mathcal L_x(z-x)+O(\delta\Psi_G^*)$.
Choosing the direction with the smaller $\Eval$ value therefore gives
$3\mathcal L_x(d)\leq\mathcal L_x(z-x)+O(\delta\Psi_G^*)$.
Since the minimum of $\mathcal L_x(z-x)$ over $\P_G$ is nonpositive,
this proves item~2 for a sufficiently large absolute constant $c$.
The running time follows from
\cref{lem:cut-partition,lem:entropy-eval,lem:entropy-internal,lem:entropy-crossing}.
\end{proof}

\subsubsection{Full algorithm and bounds}
\label{sec:decremental-potential}
When the final $\delta$ satisfies the accuracy requirement above,
apply the rebuilding procedure of \cref{sec:decremental-robustness}
at accuracy $\eps/5$.
At each rebuild, use \cref{thm:dp-mwm} to compute a $3/4$-MWM of weight $L$,
then run \cref{alg:phased-optimization} as in \cref{lem:decremental-rebuild}
at accuracy $\Theta(\lambda\eps^2)$ with a sufficiently small constant,
using the oracles above.
Maintain the integral output at accuracy $\eps/5$
using \cite[Corollary~4.19]{Chen26}.

\begin{proof}[Proof of \cref{thm:decremental}]
If the final $\delta$ fails the accuracy condition of \cref{lem:cst-concave-matching},
then $n=\otilde(\eps^{-6})$.
In this range, recompute a $(1-\eps/2)$-MWM after every deletion
using \cref{thm:dp-mwm}.
The total time is
$O(m^2\eps^{-1}\log(1/\eps))
=\otilde(n\eps^{-19})$,
within the stated bound.
Otherwise, every phase satisfies the accuracy condition.

The entropy bound in \cref{sec:decremental-robustness}
gives $\Psi_G^*\leq2\mu_w(G)$,
so $\Psi_G^*/4\leq L\leq\Psi_G^*$.
By \cref{lem:decremental-rebuild,lem:decremental-step,lem:entropy-eval,lem:cst-decremental},
$O(\eps^{-2}\log^2 n)$ rebuilds
maintain a fractional $(1-\eps/5)$-MWM.

Each phase uses $O(\delta^{-2}\log W)$ iterations
by \cref{lem:decremental-rebuild}, and the phase costs sum geometrically.
The oracle costs in \cref{lem:decremental-step,lem:entropy-eval}
therefore give
$O((m\delta^{-2}+n\delta^{-15})\log^{O(1)}n)$ time per rebuild.
Substituting $\delta=\Theta(\lambda\eps^2)$
and multiplying by the number of rebuilds gives total optimization time
$O((m\eps^{-8}+n\eps^{-47})\log^{O(1)}n)$.

For integral rounding, clear the old fractional vector
before installing each rebuilt vector.
This preserves feasibility and uses $O(m)$ coordinate updates per rebuild.
There are therefore $O(m\eps^{-2}\log^2 n)$ coordinate updates
over all rebuilds and deletions.
The rounding algorithm of \cite{Chen26}
takes $O(\eps^{-6}\log^{O(1)}n)$ amortized time per coordinate update,
adding $O(m\eps^{-8}\log^{O(1)}n)$ total time.
In this accuracy range, the number of oracle calls is polynomial in $n$.
Using fresh randomness in each call and a union bound
gives the high-probability guarantee against adaptive deletions.
\end{proof}

\section*{Acknowledgments}
The author thanks Ta-Wei Tu and Aaron Sidford for their feedback
on the introduction and acknowledges support from
the Dantzig-Lieberman Operations Research Fellowship.

\section*{AI Methodology}
The author began this project to simplify and improve
the primal-dual framework of \cite{BernsteinC26},
which uses fixed-size additive dual adjustments.
During extensive research discussions with ChatGPT 5.6 Sol and ChatGPT 6 Astra,
the models proposed an alternative strategy for dual updates.
The author connected this strategy to MWU
and to dual entropy regularization for packing linear programs,
and reformulated the algorithm
under a general first-order optimization framework.

To implement the framework efficiently,
the models proposed the cut-partition construction
that avoids computing a Gomory--Hu tree (\cref{sec:cut-construction}),
and adapted the matching-composition argument of \cite{Chen26}
to speed up our fully dynamic reduction by a factor of $1/\eps$
(\cref{subsec:dynamic-temporal-composition}).
Additionally, the models helped extend the framework to concave objectives
(\cref{sec:decremental-rebuild}).
The author verified the correctness and originality of all content, including references.
\bibliographystyle{alpha}
\bibliography{reference}

\end{document}